\documentclass{article}

\usepackage[utf8]{inputenc} 
\usepackage[T1]{fontenc}    
\usepackage{hyperref}       
\usepackage{url}            
\usepackage{booktabs}       
\usepackage{amsfonts}       
\usepackage{nicefrac}       
\usepackage{microtype}      
\usepackage{xcolor}         
\usepackage{amsmath}
\usepackage{xspace}
\usepackage{cleveref}
\usepackage{algorithm}
\usepackage{algpseudocode}
\usepackage{graphicx} 
\usepackage{indentfirst}
\usepackage{subcaption}
\usepackage{todonotes}
\usepackage{pgfplots}
\usepackage{colortbl}
\usepackage{natbib}
\graphicspath{ {Images/} }

\newcommand{\mathdefault}[1][]{}

\newcommand{\rex}{\textsc{ReX}\xspace}
\newcommand{\albert}{\textsc{Albert}\xspace}
\newcommand{\sbert}{\textsc{S-Bert}\xspace}
\newcommand{\bert}{\textsc{Bert}\xspace}
\newcommand{\rake}{\textsc{Rake}\xspace}
\newcommand{\yake}{\textsc{Yake}\xspace}
\newcommand{\clip}{\textsc{Clip}\xspace}
\newcommand{\clipscore}{\textsc{ClipScore}\xspace}
\newcommand{\yoloworld}{\textsc{YoloWorld}\xspace}
\newcommand{\llava}{\textsc{Llava}\xspace}
\newcommand{\blip}{\textsc{Blip}\xspace}
\newcommand{\opendalle}{\textsc{OpenDALLE}\xspace}
\newcommand{\textcaps}{\textsc{TextCaps}\xspace}
\newcommand{\sdxl}{\textsc{SDXL-Turbo}\xspace}
\newcommand{\sd}{\textsc{Stable-Diffusion}\xspace}
\newcommand{\kandinsky}{\textsc{Kandinsky}\xspace}

\newtheorem{lemma}{Lemma}
\newtheorem{definition}{Definition}
\newtheorem{proposition}{Proposition}

\newcommand{\lem}{\begin{lemma}}
\newcommand{\elem}{\end{lemma}}
\newcommand{\pro}{\begin{proposition}}
\newcommand{\epro}{\end{proposition}}
\newcommand{\dfn}{\begin{definition}}
\newcommand{\edfn}{\end{definition}}

\pgfplotsset{compat=1.18}

\crefname{equation}{equation}{equations}
\crefname{figure}{figure}{figures}
\crefname{section}{section}{sections}

\title{It's a Feature, Not a Bug!\\ Measuring Fluidity in Image Generators}
\author{
    Aditi Ramaswamy\\
    \texttt{aditi.ramaswamy@kcl.ac.uk}
    \and
    Melane Navaratnarajah\\
    \texttt{melane.navaratnarajah@kcl.ac.uk}
    \and
    Hana Chockler\\
    \texttt{hana.chockler@kcl.ac.uk}
}

\begin{document}

\maketitle

\begin{abstract}
With the rise of freely available image generators, AI-generated art has become the center of a series of heated debates, one of which concerns the concept of human creativity. Can an image generation AI exhibit ``creativity'' of the same type that artists do, and if so, how does that manifest? Our paper attempts to define and empirically measure one facet of creative behavior in AI, by conducting an experiment to quantify the ``fluidity of prompt interpretation'', or just ``fluidity'', in a series of selected popular image generators. To study fluidity, we (1) introduce a clear definition for it, (2) create chains of auto-generated prompts and images seeded with an initial "ground-truth: image, (3) measure these chains' breakage points using preexisting visual and semantic metrics, and (4) use both statistical tests and visual explanations to study these chains and determine whether the image generators used to produce them exhibit significant fluidity.
\end{abstract}

\section{Introduction}

There are many metrics and measures used to evaluate different aspects of image generators. However, one concept has been overlooked for a long time: creativity. This is because creativity is deeply difficult to define, let alone quantify. We aim to change this by introducing a new creativity measure to balance out the existing metric of faithfulness, which measures how strongly a generated image matches the textual prompt used to create it. Our proposed measure is ``fluidity of prompt interpretation''. We measure it through building a game between auto-generated captions and images, and analyzing the results through both statistical and visual means. The proposed measure has a fine granularity, allowing for comparison of different image generators by placing them on a scale of ``fluid'' to ``faithful''. To the best of our knowledge, this is the first attempt at such a measure.

\subsection*{Related Work}

Some researchers argue that creativity is derived from a ``socio-cultural context'', thus excluding AI models, which ``[lack]... feelings... or [the ability to] reflect''~\citep{Oppenlaender_2022, Kaufman_2019, Wingström_2022}. However, other definitions focus on purely behavioral requirements, such as ``domain-relevant skills'', ``creativity-relevant processes'', and ``extrinsic motivation'', all of which can be exhibited by generative AI models \citep{Amabile_1983}.

A number of ethnographic studies support this. AI artists interviewed by \cite{Wingström_2022} spoke of ``co-creativity'', the synthesis of their own human creativity and generative AI processes, and stated that glitches in generated images are signs of creativity rather than bugs. Other studies use the Alternate Uses Task (AUT) test, an established subjective measure of creativity, to compare creative behaviors exhibited by AI chatbots and humans \citep{Koivisto_2023, Haase_2023}. The results showed that human scorers are often unable to differentiate AI-generated results from human-generated ones. Similarly, recent research indicates that ``little to no'' human intervention is needed for an image generation model to produce ``high-quality'' art, as DALL-E can create complex and detailed pieces from very simple text prompts such as emojis or singular letters \citep{Oppenlaender_2022}. 

Using a process-oriented definition like that described by \cite{Amabile_1983}, and supported by the studies conducted by \cite{Wingström_2022, Koivisto_2023, Haase_2023, Oppenlaender_2022} we infer that some AI models may have the potential to exhibit creative behavior. This forms the motivation for our experiment.

\section{Background \& Experiment Construction}
Here we provide definitions for fluidity and breakage point/chain length, as well as a detailed breakdown of our experimental setup, chain construction, and breakage calculation metric. 

\subsection{Existing Image Generation Metrics}

There are a range of existing techniques for evaluating image generators which we took into consideration when creating our fluidity measure. In particular, fidelity focuses on the visual similarity between generated and ground truth images, while faithfulness evaluates the alignment between the prompt and generated images. \cite{Hu_2023}, for example, evaluates faithfulness in an image generation model $m$ by leveraging an LLM to answer questions about a text prompt $t$ given a generated image produced by $t$ using $m$. This is not dissimilar to our approach of generating images and comparing both the generated images and their captions to a predetermined ``ground truth''. However, there is no definitive measure to study the opposite of faithfulness: how much \textit{misalignment} is there between the prompt and a generated image? Knowing this misalignment level would enable a user to choose the model which best suits their end goal. Our measure uses aspects of both fidelity and faithfulness to gauge this misalignment.

\subsection{Defining and Measuring Fluidity}

While creativity as a concept cannot be measured, creative behavior can be experimented on to produce analyzable results. The view on glitches being indicators of creativity in \cite{Wingström_2022} led us to craft an experiment to evaluate image generators for one measure of creative behavior. This measure is ``fluidity of prompt interpretation'', or simply fluidity.

\begin{definition}[Fluidity]
    The relative extent of misalignment between the output of a given image generator and the semantics of its input prompt, placed on a scale from a hypothetical completely random image generator to a hypothetical completely faithful one, using the definition of faithfulness provided by \cite{Hu_2023}.
\end{definition}

We measure the fluidity of a given generator through constructing numerous alternating chains of generated images and captions, calculating when each chain strays too far from the original ``ground truth'' (a ``breaking point''), and then studying the statistical properties of the ensuing distribution of chain lengths.

\subsection{Computational Resources \& Justifications}
Due to computational limitations, we could not run each chain until its natural breaking point. We set a hard limit of 15 generated images, as that is long enough to let us tell whether a chain stays faithful to the ``ground truth'' image and caption, but short enough not to require more resources than we could afford. Our experiment therefore required 1000 chains of 15 generated images and 16 generated captions for 12 different image+caption generator combinations in \Cref{tab:individual-stats}, plus 15 generated captions for each of the control chain combinations shown in \Cref{tab:control-stats}. For each combination, we used 1-2 A100 GPUs. A chain length of 15 is therefore synonymous with the chain staying unbroken.

\subsection{Chain Construction}

As in Chinese Whispers, each chain in our experiment begins with a ``ground truth'', or seed image, pulled from \texttt{coco\_1000}, a set of 1000 images which we compiled from the publicly-available, non-copyrighted COCO dataset \citep{Lin_2014}. Our compilation process involved randomly selecting photographs from COCO, while using YOLO object detection to ensure that each chosen photo had a clear subject and did not prominently feature human faces \citep{Jocher_2023}. The latter criterion is because facial features are inherently so variable that determining an easily understandable breakage point between two facial images is too subjective.

To build a chain, we caption the initial ``ground truth'' seed image using one of three caption generation models from Hugging Face\footnote{\url{https://huggingface.co/}}: \textcaps, \blip, and \llava \citep{Wang_2022, Li_2022, Liu_2023}. This caption is then fed into one of four open-source image generation models: \opendalle, \kandinsky 2.2, \sd, and \sdxl \citep{Izquierdo_2024, Razzhigaev_2023, Rombach_2022, Sauer_2023}. To ensure determinism in our experiment, we seeded each image generator with a constant value (the caption generators are already deterministic). We also tested reproducibility by running miniature versions of our experiments for each image generator (using \llava) and used Mann-Whitney U tests to confirm that the chain length frequency distributions produced by each of the new experiments had no statistically significant difference from those we derived from the main experiment.

For \sd, we added negative and positive prompt lists to ensure that the outputs to remain photo-like so they could be better compared to the ``ground truth'' photo\citep{Berger_2023}. We also set the guidance scale (when it was a viable parameter) to the maximum value, as we wanted to observe the fluidity of these image generation models when they are explicitly instructed to adhere to a given text prompt as much as possible.

Our reason for choosing multiple captioning models is to mitigate the effect a single captioning tool may have on the effects: if it produces inaccurate captions, it could cause the chains to break early even if the image generator is faithful in interpreting a given prompt. Using three different captioning tools balances this out by allowing us to compare the three to determine whether the captioning tool has a significant influence on chain breakage. Since image generation models tend to do best with concise prompts, we limited the maximum caption length to 50 for \textcaps, and for \llava we requested a ``concise caption'' within the parameters. \blip used default parameters, but produced captions of roughly the same length as \llava and \textcaps.

We also created a control group of ``dummy'' chains to represent the products of running our experiment with a hypothetically maximally faithful, minimally fluid image generator. We determined that such a generator would produce extremely similar images at each step in the chain, sticking to the original image and prompt with extreme fidelity and faithfulness. We pulled 15 images each of ``dummy'' images from a Kaggle dataset of bears, zebras, and giraffes, common subjects in our input image dataset \citep{Likhon_2024}. To simulate the 1000 image chains produced by each of our true experiments, we randomly shuffled each set of 15 images of the same category 333 times, captioned each image as if it were a real chain, and then calculated the breaking metrics using the same formulae as the real chains. The creation of this control group was necessary in order to have a null hypothesis to which the image generators could be compared: because the images in each control chain are extremely similar, they represent the most faithful interpretations of the seed prompt, thus simulating an image generator which always acts with extremely low fluidity and extremely high faithfulness.

We check whether our experimental results are statistically significant by using the Mann-Whitney U non-parametric test to compare the chain length frequency distributions of chain lengths for each image and caption generator combination with a corresponding control group distribution, and we compare fluidity levels amongst different image and caption generator combinations by using Kullback-Leibler divergence.

\subsection{Image Guidance, Glitches, \& Breaking Calculations}

Our breakage criteria depend on the presence of generative AI glitches, caused by semantic misunderstandings between the text prompts and the embeddings of the generative models used to produce images \citep{Chefer_2023}. Often, these can result in the intended subjects from the text prompt either not being generated (``catastrophic neglect''), or being given less importance than background information from the text prompt (``incorrect attribute binding''). Examples from \cite{Chefer_2023} include the prompt ``a yellow bowl and a blue cat'' producing blue-and-yellow bowls, and ``a yellow bow and a brown bench'' producing the specified items, but coloring both yellow. Computational creativity also views ``incongruities'' and ``anomalies'' such as these as \emph{opportunities for further search}, or paths which may lead to \emph{``surprisingly meaningful results''}~\citep{Veale_2019}. This fluidity is an important aspect of creativity: humans often interpret media, such as literary works, in unexpected or unusual ways. Therefore, examining unusual semantic interpretations performed by image generators through the lens of creative, rather than buggy, behavior, can lead to new insights on the contrasts and similarities between humans' and machines' understanding of the world.

\begin{definition}[Breaking Point/Chain Length]
    The iteration $x$, within a given chain $c$, wherein $img_x$, the generated image at $x$, is too distant from the seed image's caption $s_{cap}$ to be considered faithful to the semantic information contained in $s_{cap}$. This distance is measured through \Cref{fig:breakage-algorithm}, which utilizes a series of preexisting metrics to determine the distance from $img_x$ to the seed image.
\end{definition}

To measure chain length, we take into account multiple facets of each step in the chain, including objects found within the current generated image as well as the semantics of the current generated caption. These facets are compared to the initial seed image/caption pair using a number of metrics. Firstly, to quantify the disparities between the initial caption and each subsequent caption in the chain, we use variations of BERT to measure semantic change. Secondly, we use both \clip and YOLO to quantify the disparities between the initial seed image and each subsequent image in the chain, by extracting the likeliest labels from each image and comparing them. Our intent in using multiple semantic scorers and object detectors is to try to mitigate any skew one particular tool may introduce into the results.

For semantic scoring, we use the \albert and \sbert variations of the original \bert model \citep{Lan_2019, Reimers_2019}. For \albert, we feed in labels representing the semantic meaning of each caption being compared: the original seed one and the caption at the current step of the chain. These labels are generated using the unsupervised keyword extraction algorithm \yake, and if that yields no results, a second attempt at extracting labels is made using \rake \citep{Campos_2020, Rose_2010}. Both \yake and \rake are fast, document length-agnostic, and can extract keywords from a single document, making them ideal choices for keyword extraction from captions, which are short documents that do not belong to a larger corpus. \albert then uses cosine similarity to calculate the distance between the two sets of labels. \sbert takes in the raw text for each caption, and turns it into a word embedding representation. These two embeddings are also compared using cosine similarity. Our semantic scoring metric considers the chain ``broken'' if the result of both metrics falls below 0.5, or a less than 50\% similarity to the original caption. Example values for these metrics can be found in \Cref{fig:CaptionScores}.

To justify this threshold, we ran a few experiments on a sample of 100 images from our \texttt{coco\_1000} dataset. We determined that 0.25 is too low to produce meaningful results as almost all the chains remain unbroken, while 0.75 is too high, causing most chains to break at the first iteration.

For measuring differences in the subjects of the images, we use two object detection architectures, \clip and \yoloworld \citep{Radford_2021, Cheng_2024}. Torchmetrics' \clipscore is used to calculate the similarity between the current image and the initial caption, wherein a larger value means a greater difference. A threshold of 20, based on smaller experiments run beforehand, was used to determine breakage based on \clipscore. Additional breakage metrics based on the images themselves were implemented using the class labels provided by both \clip and \yoloworld. To compare similarity, we use algorithm $LABEL\_SIM(curr\_img\_labels, init\_img\_labels)$, as defined in \Cref{fig:label-sim}.

Examples of object detector breakage results can be found in \Cref{fig:LabelScores}.

For each step $x$ of 15 in each chain, the breakage algorithm \Cref{fig:breakage-algorithm} is called in order to compare the image and caption generated at step $x$ with the ``ground truth'' image and its caption which were determined at the very beginning of the chain. 

Comprehensive examples of chain breakages can be found in \Cref{fig:dalle-llava-appendix}, \Cref{fig:kand-llava-appendix}, \Cref{fig:sd-llava-appendix}, and \Cref{fig:sdxl-llava-appendix} in the appendix. 

\subsection{Analysis Tools}

In order to understand the underlying reasons as to why some chains broke, we used \rex, an improved re-implementation of \textsc{DeepCover} which provides visual explanations for the pre-determined ResNet152 labels applied to an input image \citep{He_2016, Chockler_2022}.

\section{Results \& Analysis}
\subsection{Statistical Results} 

\begin{figure}
\centering
\begin{subfigure}{0.4\textwidth}
    \centering
    \includegraphics[width=1\textwidth]{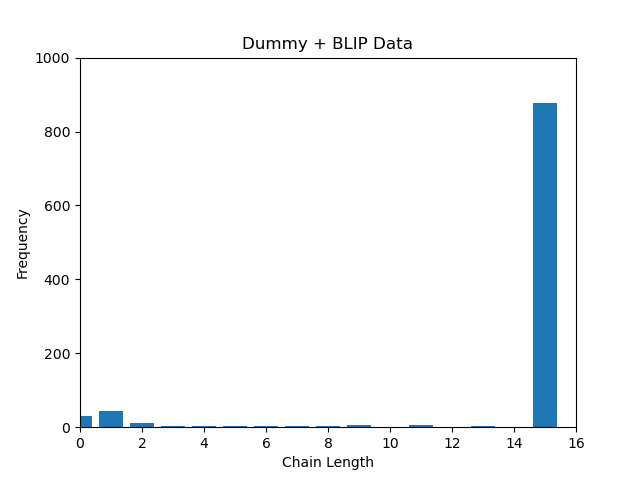}
    \caption{CONTROL + \blip}
    \label{fig:cblip}
\end{subfigure}
\hfill
\begin{subfigure}{0.4\textwidth}
    \includegraphics[width=1\textwidth]{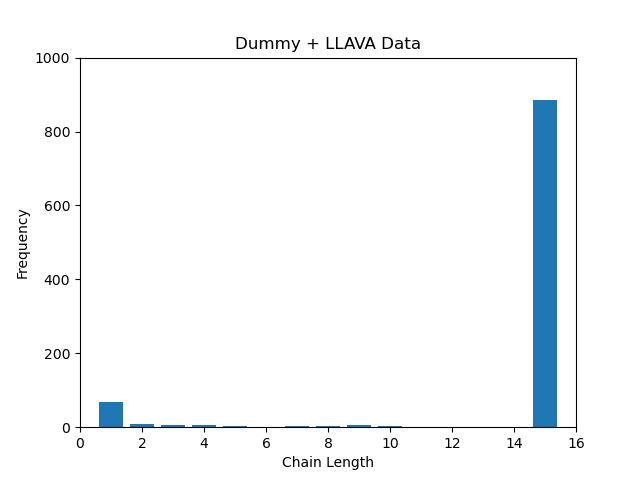}
    \caption{CONTROL + \llava}
    \label{fig:cllava}
\end{subfigure}
\hfill
\begin{subfigure}{0.4\textwidth}
    \includegraphics[width=1\textwidth]{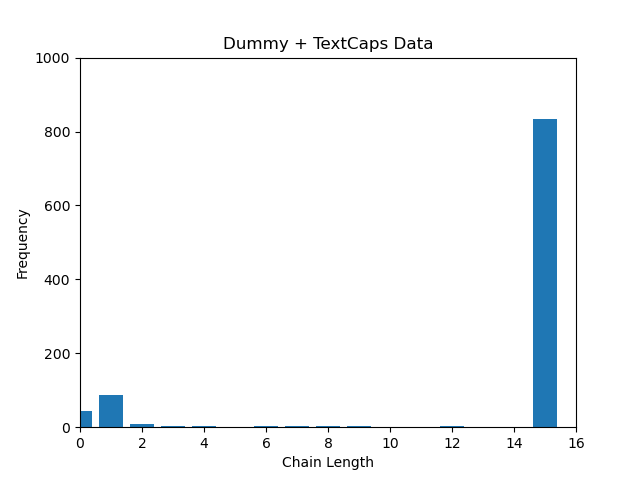}
    \caption{CONTROL + \textcaps}
    \label{fig:ctc}
\end{subfigure}
\caption{The chain length frequency distributions for all three control group combinations.}
\label{fig:dummy-graphs}
\end{figure}

To formalize our conjecture about fluidity and chain breakage, we established the following null hypothesis: \textit{for any given combination of image generator and caption generator, the frequency distribution of chain lengths would not show a statistically significant difference from the frequency distribution of chain lengths in the control group produced with corresponding captioning tool}. The chain length frequency distributions for the control group data are shown in \Cref{fig:dummy-graphs}.

When the frequencies of chain lengths/breakage points for each combination of image generator and caption generator were plotted, they were all negatively skewed according to the Shapiro-Wilk statistical test \citep{Shapiro_1965}. Therefore, we chose to use the two-sided Mann-Whitney U non-parametric test to compare the chain data to the control group data \citep{Mann_1947}. We used the implementation from the \texttt{scipy stats} library to perform both this test and KL divergence \citep{Virtanen_2020}.

\begin{figure}[!htb]
    \centering
    \input{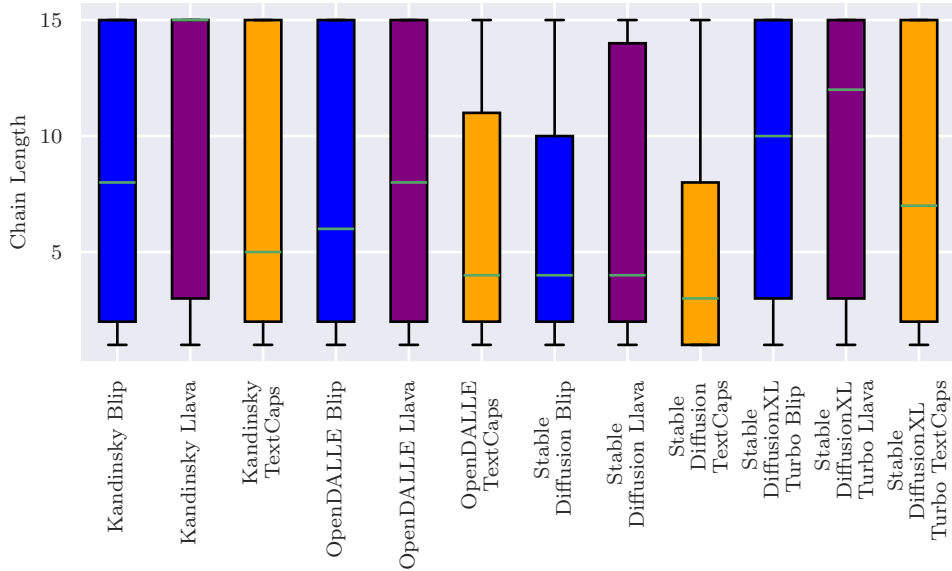}
    \caption{A box plot of the distribution of chain lengths for the different combinations of models}
    \label{fig:3}
\end{figure}

We performed the Mann-Whitney U test with a default p-value threshold of 0.05 on the frequency data compiled from the \opendalle, \kandinsky, \sd, and \sdxl chain outputs and the control group data corresponding to each of the caption generators \blip, \llava, and \textcaps. We also used the Mann-Whitney U test to compare image generators and captioning tools amongst themselves. In all, we performed the Mann-Whitney U test on 45 different pairs of images (see \Cref{tab:control-comparisons}, \Cref{tab:img-comparisons}, and \Cref{tab:cap-comparisons}), so we used the Bonferroni correction by dividing 0.05 by 45, leading to a p-value significance threshold of 0.0011 \citep{Sedgwick_2012}.

As shown in \Cref{tab:control-comparisons}, most of the combinations' chain length frequency distributions differed from the expected behavior of the corresponding control groups with p-values below 0.0011. The exceptions were \kandinsky-\blip, \kandinsky-\textcaps, and \sdxl-\textcaps.

From this, we determine that \sd and \opendalle both unequivocally reject the null hypothesis that the chain length frequency distribution derived from running our experiment using a particular image generation model will not differ statistically significantly from the chain length frequency distribution derived from running our experiment using control chain data. \kandinsky and \sdxl are ambiguous, as some of the captioning models pushed their p-values very slightly above the 0.0011 threshold while other captioning models had p-values slightly below the threshold. However, as shown in Table \ref{tab:cap-comparisons}, none of the Mann-Whitney tests conducted on pairs which use the same image data and different captioning models had p-values close to the 0.0011 threshold. These results indicate that while some captioning tools may influence results slightly, the chosen captioning tool alone does not have as strong an impact on chain length frequency distributions as does the chosen image generator.

\begin{figure}[!htb]
    \centering
    \includegraphics[scale=0.3]{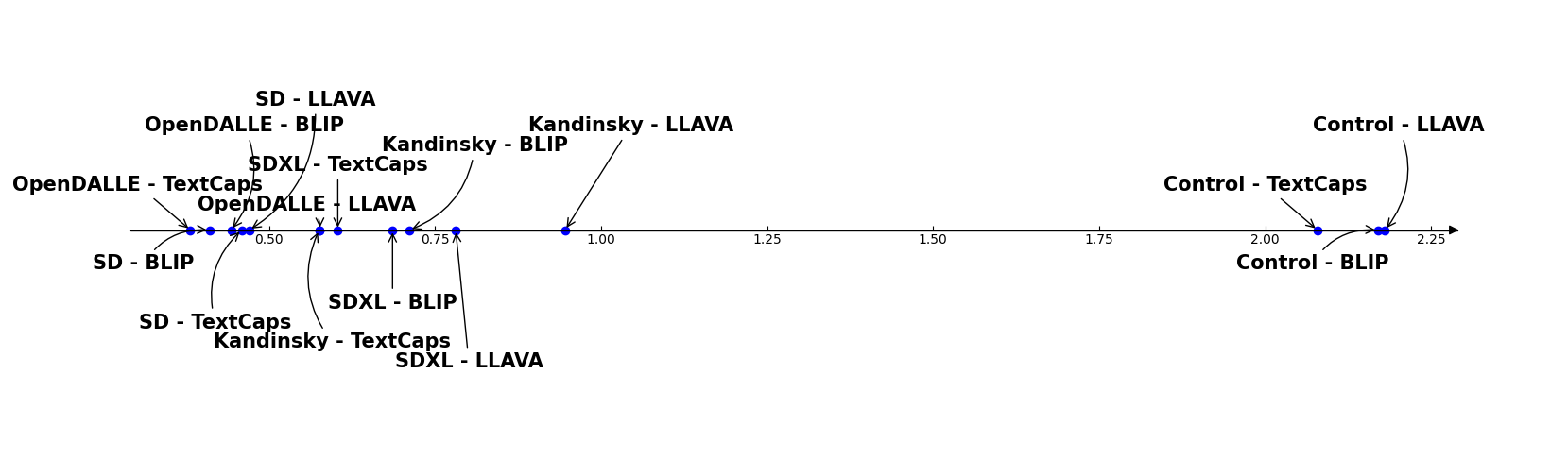}
    \caption{The different combinations of models plotted on a scale of fluidity (the direction is fluid $\rightarrow$ faithful, with higher values representing more faithfulness)
    using the Kullback-Leibler divergence.}
    \label{fig:FluidityScale}
\end{figure}

Although the Mann-Whitney U results showed no statistically significant differences between the distributions produced by the four different image generators in comparison to each other (see \Cref{tab:img-comparisons}), the fact that a few of the image-caption generator combinations did not reject the null hypothesis meant that subtle differences did exist between different image generators. Why, for example, does \sd-\textcaps show statistically significant fluidity, while \sdxl-\textcaps does not? We decided to build a quantitative scale of fluidity to faithfulness for the various combinations. To do this, we used Kullback-Leibler divergence (KL) to calculate the distance between the chain length frequency distribution for each combination and the uniform distribution. We selected the uniform distribution using the reasoning that a hypothetical image generator $H_i$ which does not rely on semantic information at all would generate random images, representing a maximally fluid state. This unpredictability could be simulated using a uniform distribution.  \Cref{fig:FluidityScale} shows the KL scores for each combination when compared with the uniform distribution.

The further a combination is from the maximally fluid state, the higher its KL score will be. As can be seen here, \opendalle + \textcaps is the closest to maximally fluid, with a KL score of 0.38, while Control + \llava is the furthest from maximally fluid, with a KL score of 2.18.

\subsection{Analysis}

\begin{figure}[!htb]
    \begin{subfigure}{\linewidth}
    \centering
    \includegraphics[width=1.5cm]{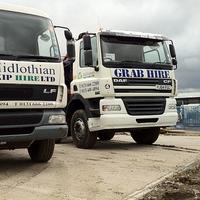}
    \includegraphics[width=1.5cm]{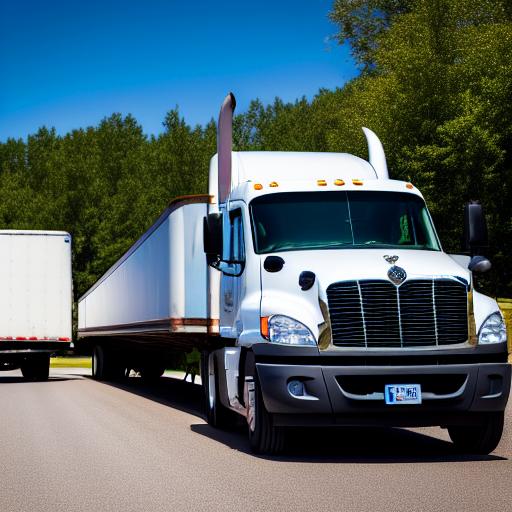}
    \includegraphics[width=1.5cm]{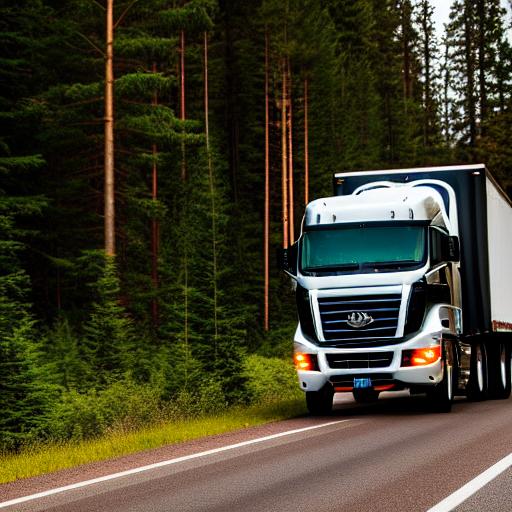}
    \includegraphics[width=1.5cm]{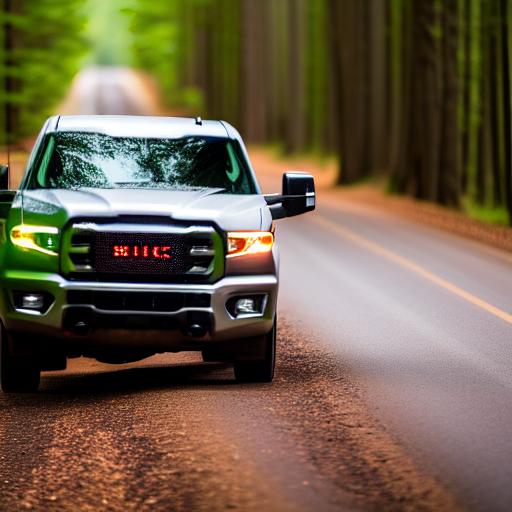}
    \includegraphics[width=1.5cm]{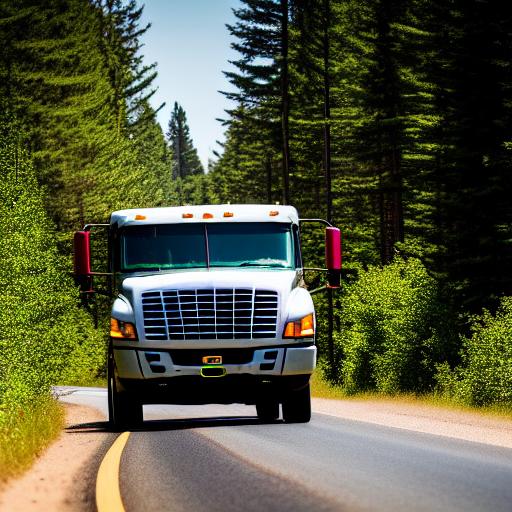}
    \includegraphics[width=1.5cm]{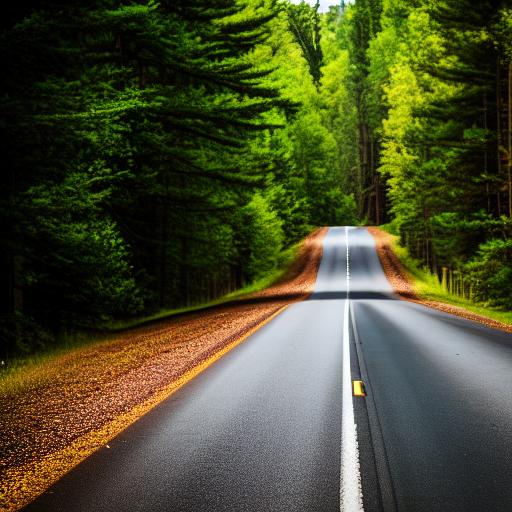}
    \includegraphics[width=1.5cm]{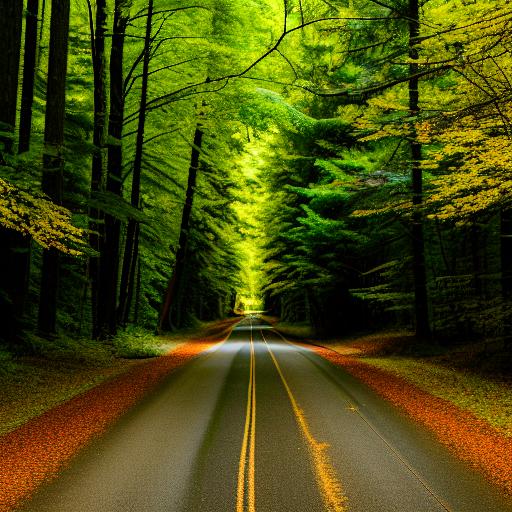}
    \includegraphics[width=1.5cm]{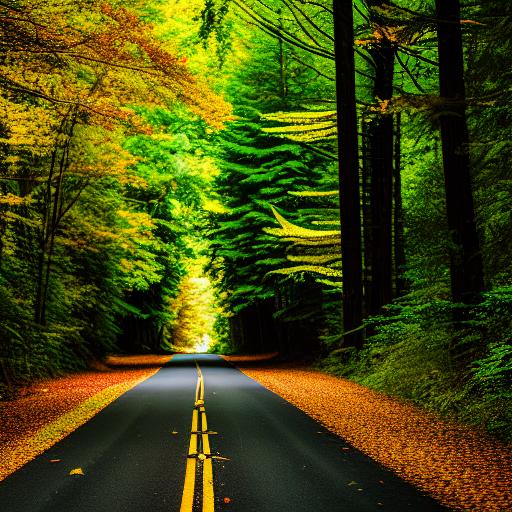}
    \includegraphics[width=1.5cm]{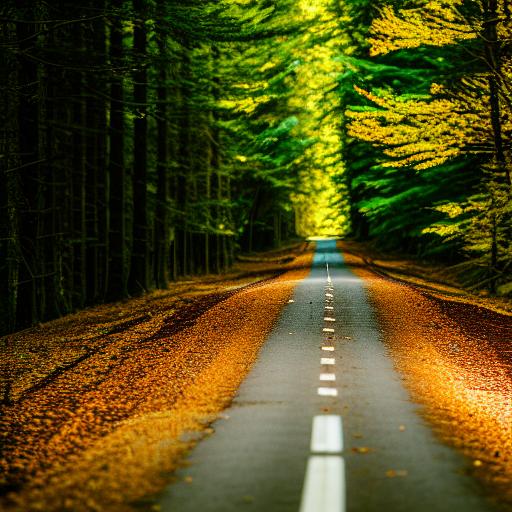}
    \includegraphics[width=1.5cm]{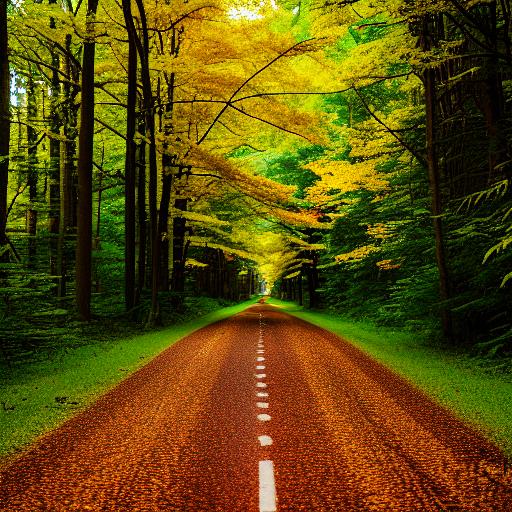}
    \includegraphics[width=1.5cm]{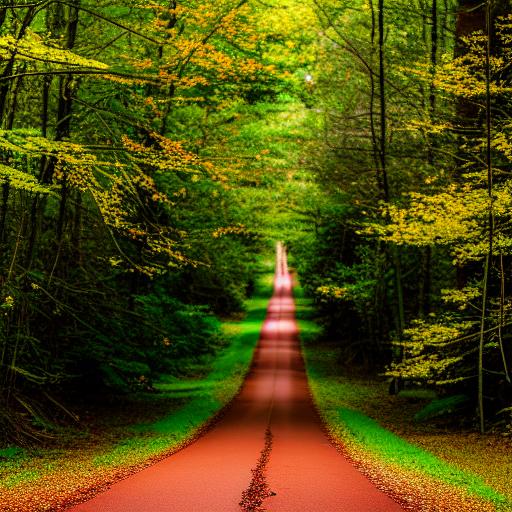}
    \includegraphics[width=1.5cm]{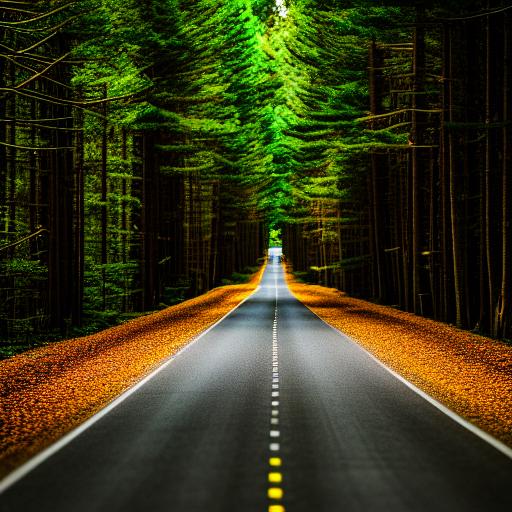}
    \includegraphics[width=1.5cm]{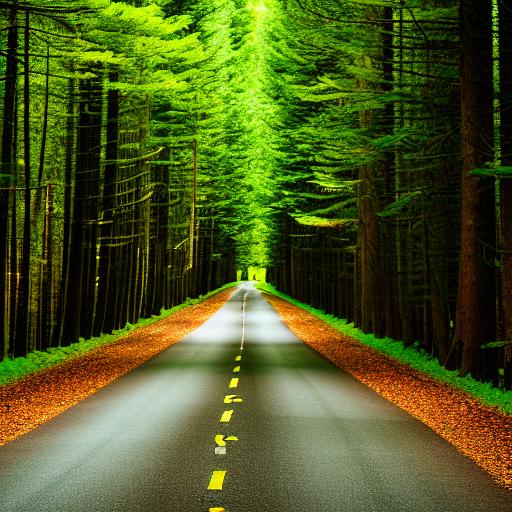}
    \includegraphics[width=1.5cm]{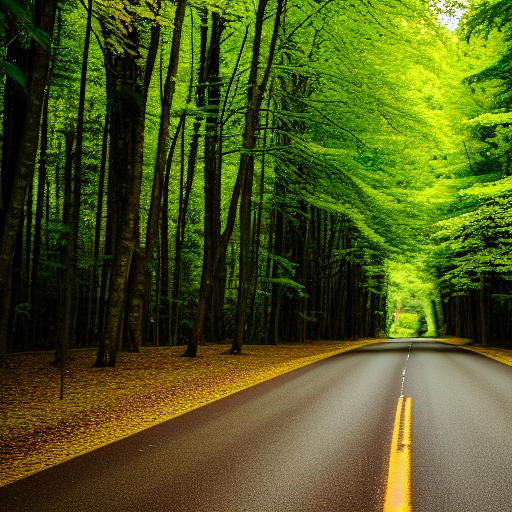}
    \includegraphics[width=1.5cm]{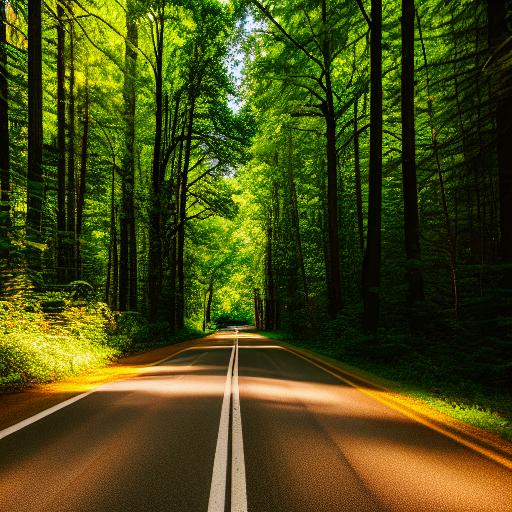}
    \includegraphics[width=1.5cm]{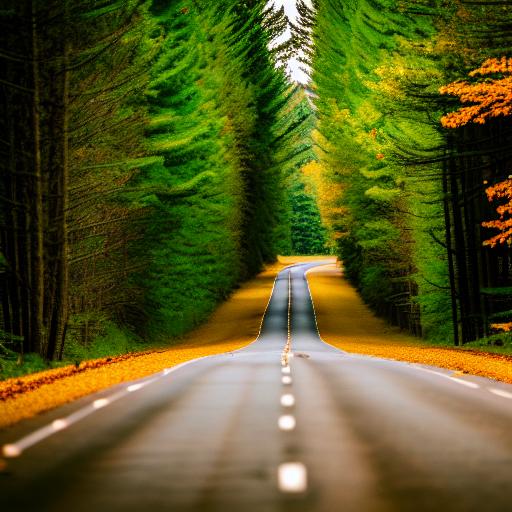}
    \caption{Chain produced using \sd and \blip.}
    \end{subfigure}\par\medskip
    \begin{subfigure}{\linewidth}
    \centering
    \includegraphics[width=1.5cm]{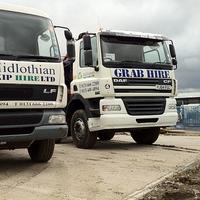}
    \includegraphics[width=1.5cm]{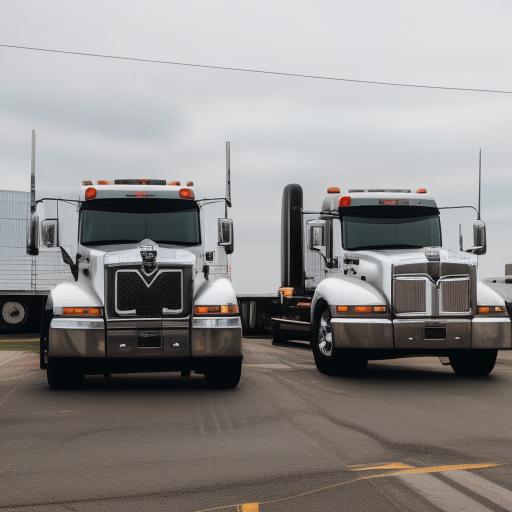}
    \includegraphics[width=1.5cm]{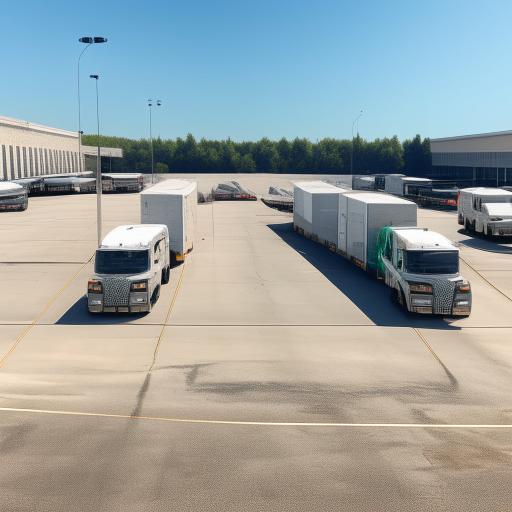}
    \includegraphics[width=1.5cm]{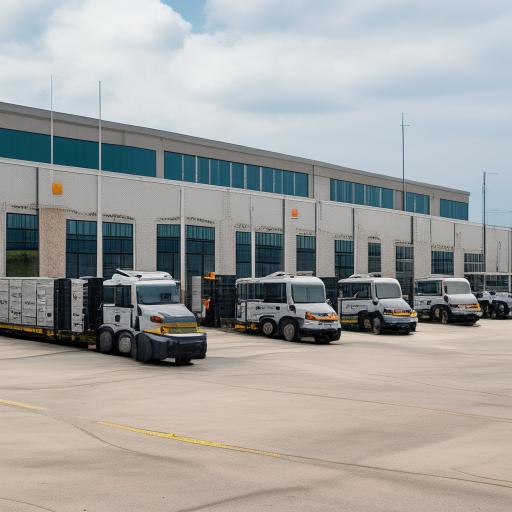}
    \includegraphics[width=1.5cm]{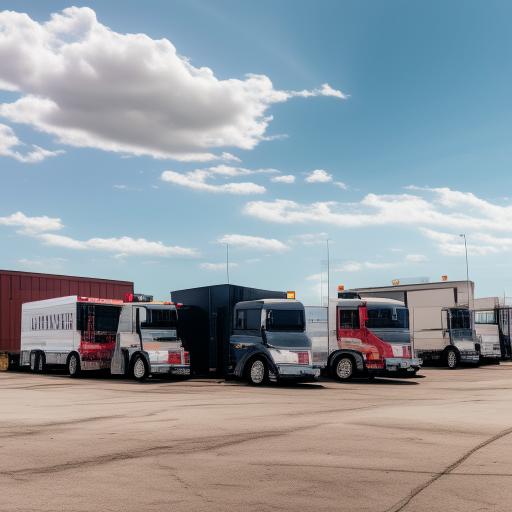}
    \includegraphics[width=1.5cm]{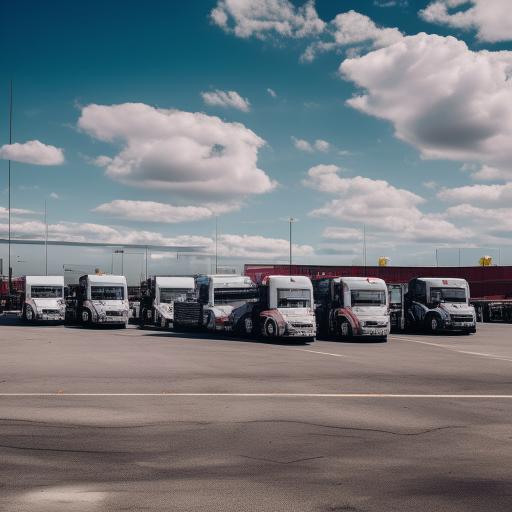}
    \includegraphics[width=1.5cm]{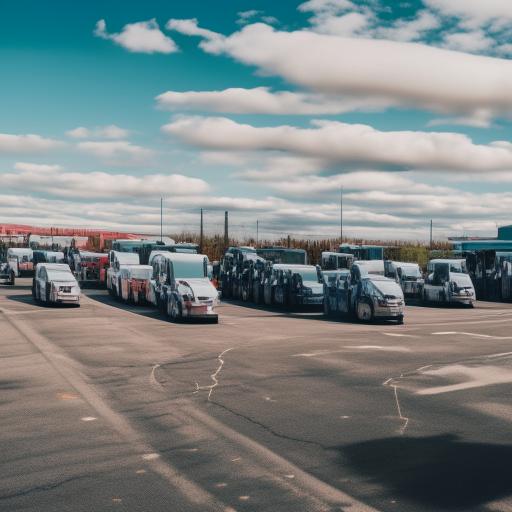}
    \includegraphics[width=1.5cm]{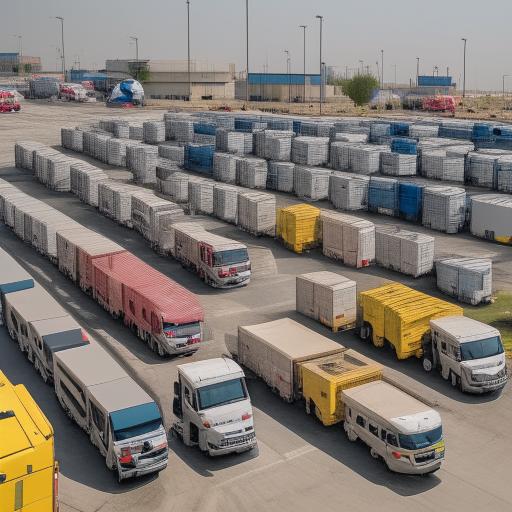}
    \includegraphics[width=1.5cm]{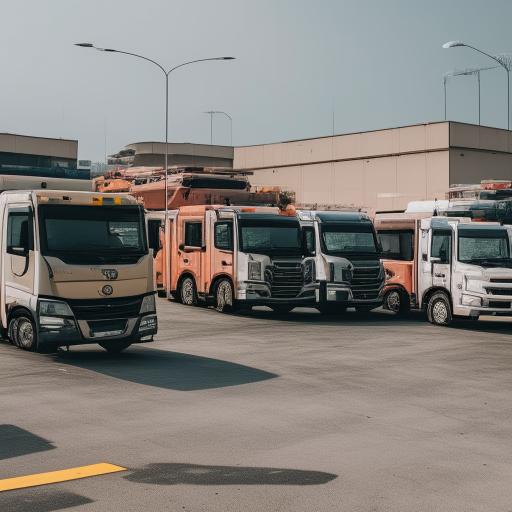}
    \includegraphics[width=1.5cm]{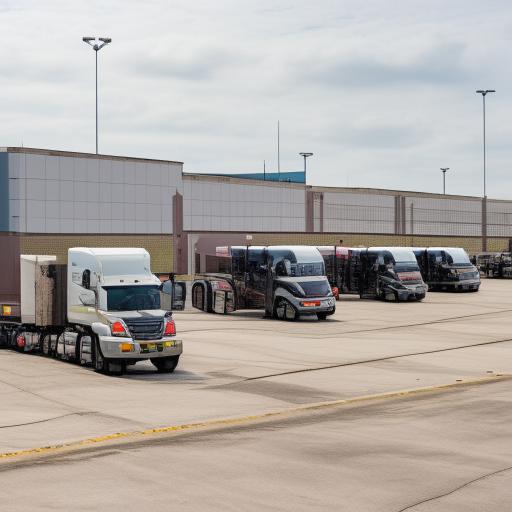}
    \includegraphics[width=1.5cm]{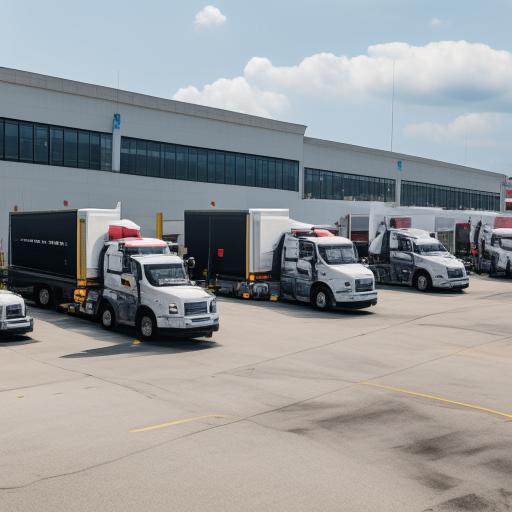}
    \includegraphics[width=1.5cm]{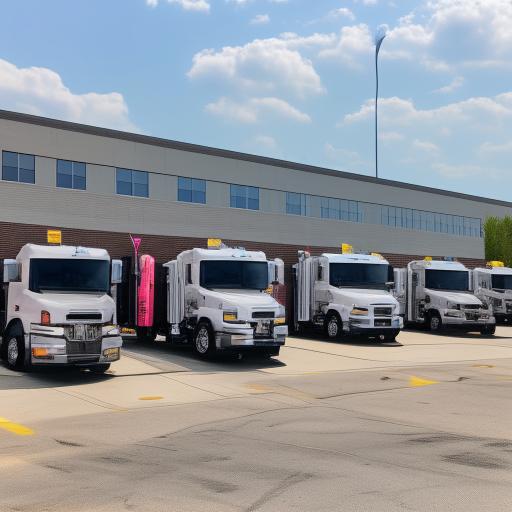}
    \includegraphics[width=1.5cm]{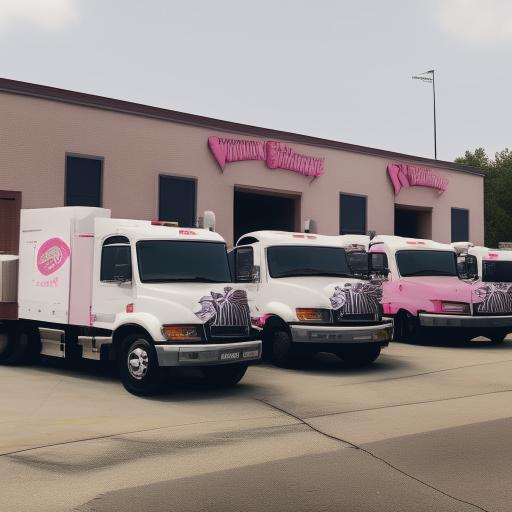}
    \includegraphics[width=1.5cm]{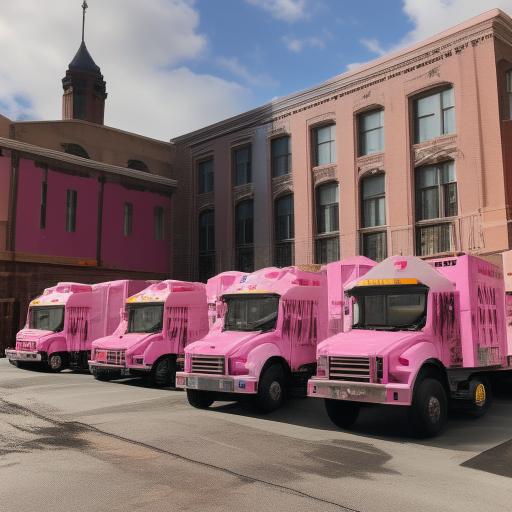}
    \includegraphics[width=1.5cm]{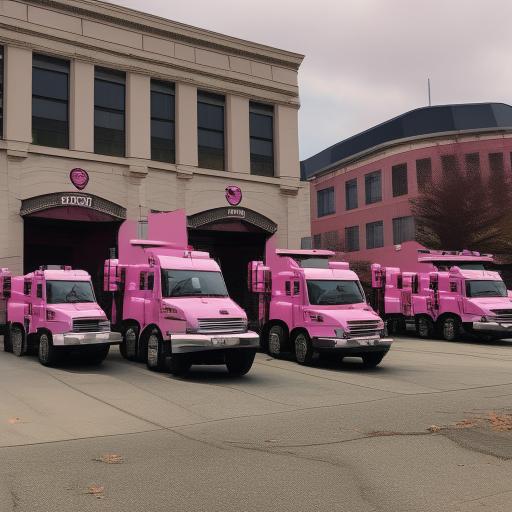}
    \includegraphics[width=1.5cm]{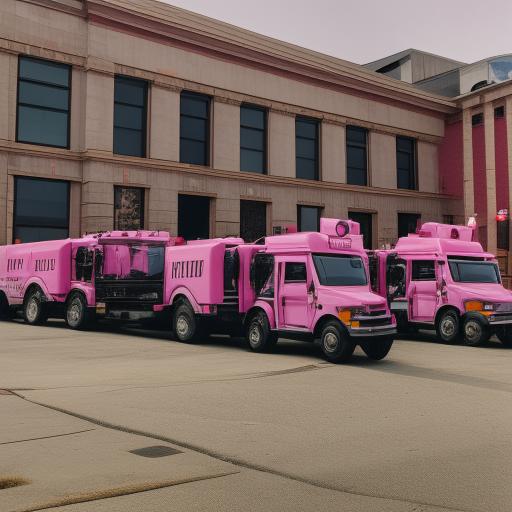}
    \caption{Chain produced using \kandinsky and \blip.}
    \end{subfigure}
    \caption{Example chains produced by ``ground truth'' image 0011, labeled ``truck'', where the top left is the ``ground truth'' image and the rest are generated.}
    \label{fig:chains-0011}
\end{figure}

Visual analysis of the outputs is integral to understanding how and why fluidity of prompt interpretation manifests within a given chain. Some chains are easy to understand at first glance, while others require more algorithmic analysis. As an example of the former case, \Cref{fig:chains-0011} shows two chains seeded with the same image, 0011, from the ``ground truth'' dataset. Both chains were created using \blip for captioning, but one used the \kandinsky model for image generation, while the other used \sd. On the scale described in \Cref{fig:FluidityScale}, \sd + \blip has a KL score of 0.41 and is more fluid than \kandinsky + \blip, which has a KL score of 0.71.

The initial caption for the ``ground truth'' image, 0011, was ``two large trucks parked next to each other on a road''. As seen in the \sd chain, the images quickly diverged from focusing on trucks to the surrounding scenery, with the last generated image in the chain being captioned ``a view of a road with trees lining both sides of it''. On the other hand, the \kandinsky chain continues to feature trucks as the main focus, with the final generated image in the chain being captioned ``several pink trucks are parked in front of a building''. The \sd chain broke at iteration 6 due to low \clip and \yoloworld label similarities between the ``ground truth'' seed image and the fifth generated image, while the \kandinsky chain remained unbroken. The reasons for these chain lengths are clear: \sd strayed away from the original prompt by interpreting its intended focus as the road, adding the forest element spontaneously, while the \kandinsky chain continued to interpret the intended focus of the caption as the trucks, producing images which are relatively faithful to the original seed image despite small changes such as the color of the vehicles.

\begin{figure}[!htb]
    \begin{subfigure}{\linewidth}
    \centering
    \includegraphics[width=1.5cm]{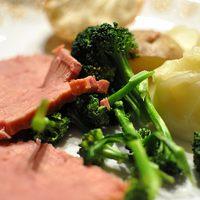}
    \includegraphics[width=1.5cm]{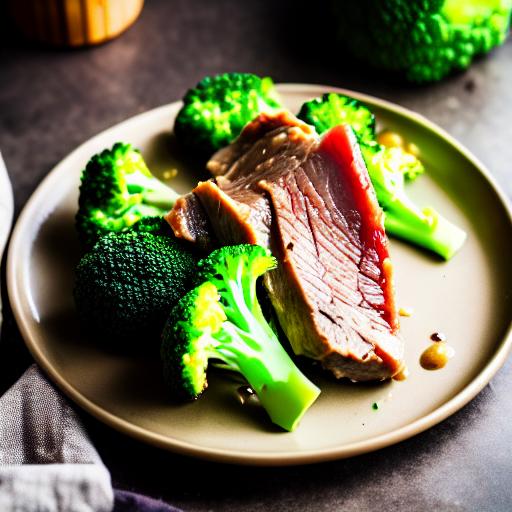}
    \includegraphics[width=1.5cm]{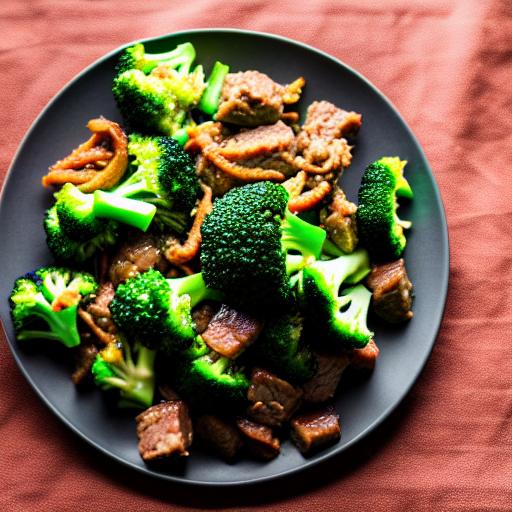}
    \includegraphics[width=1.5cm]{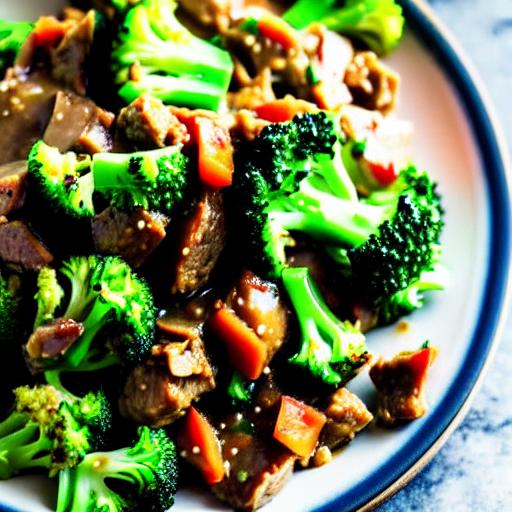}
    \includegraphics[width=1.5cm]{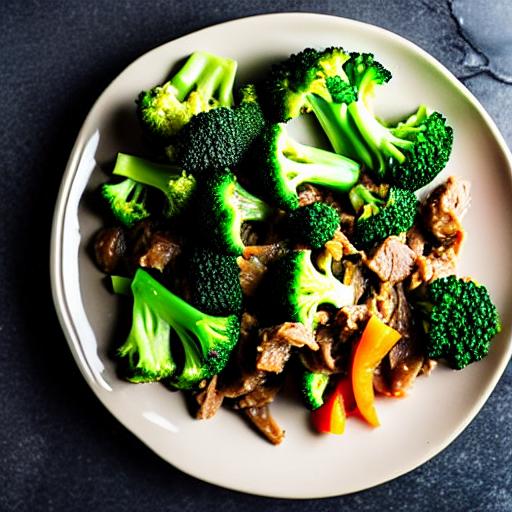}
    \includegraphics[width=1.5cm]{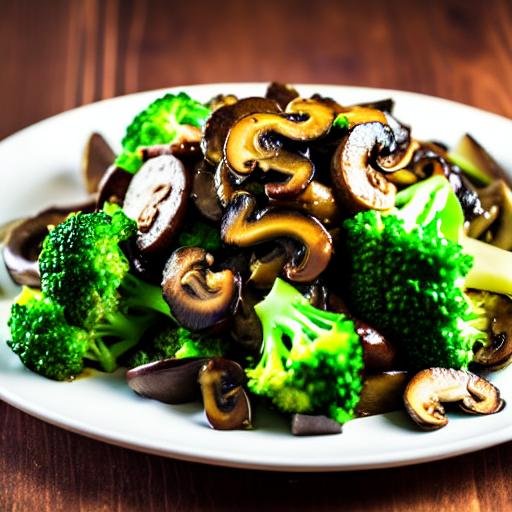}
    \includegraphics[width=1.5cm]{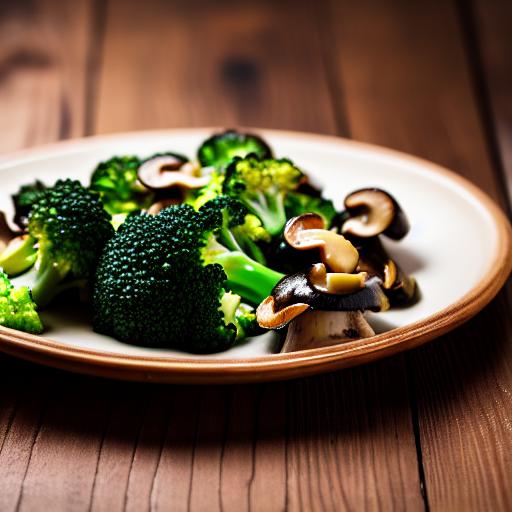}
    \includegraphics[width=1.5cm]{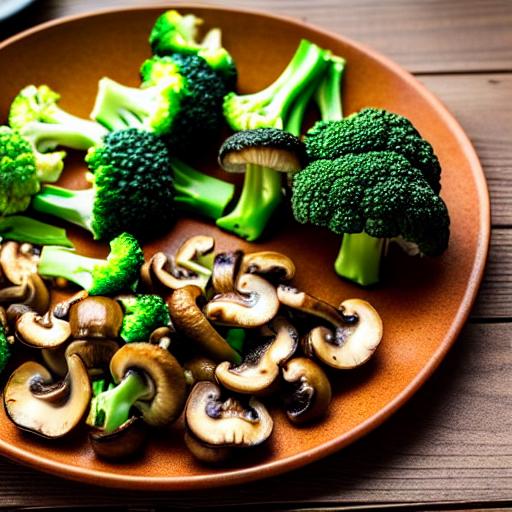}
    \includegraphics[width=1.5cm]{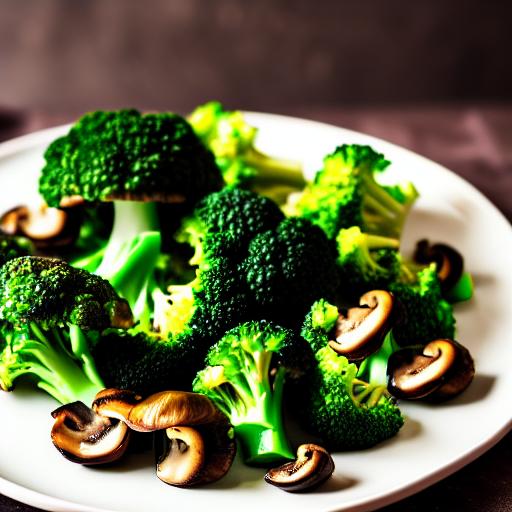}
    \includegraphics[width=1.5cm]{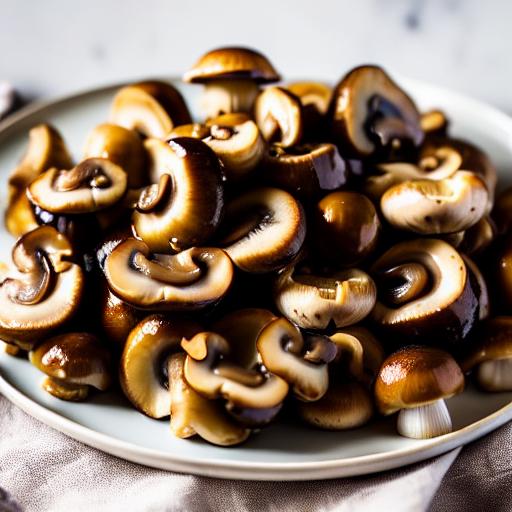}
    \includegraphics[width=1.5cm]{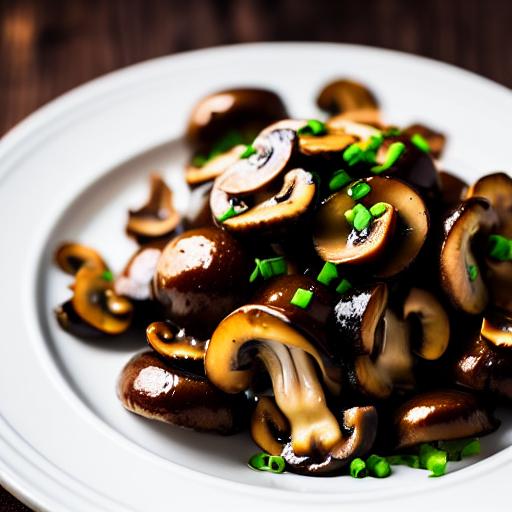}
    \includegraphics[width=1.5cm]{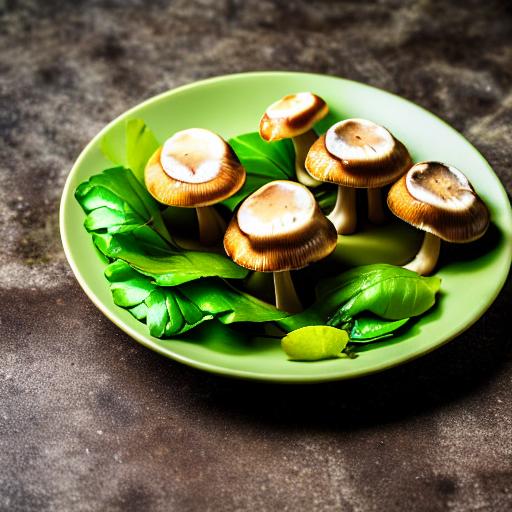}
    \includegraphics[width=1.5cm]{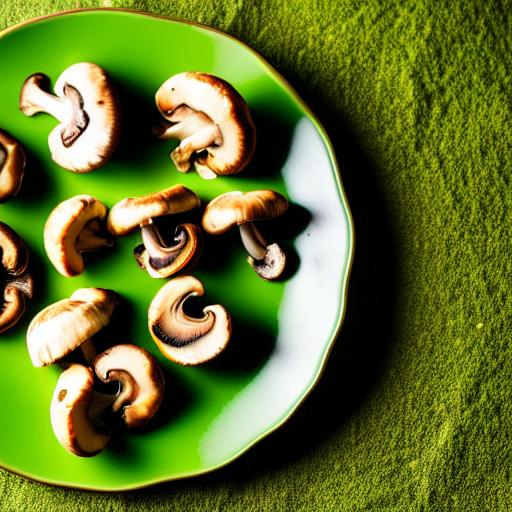}
    \includegraphics[width=1.5cm]{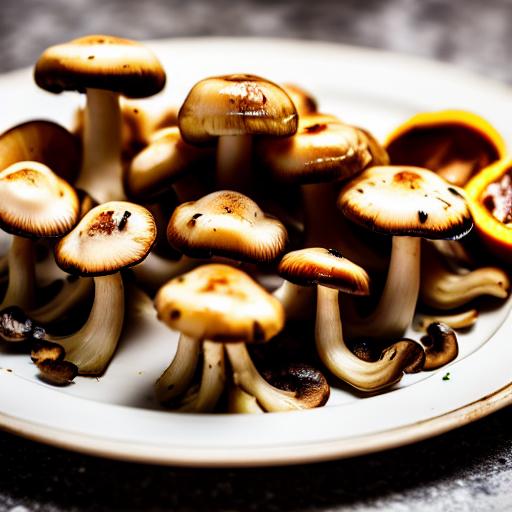}
    \includegraphics[width=1.5cm]{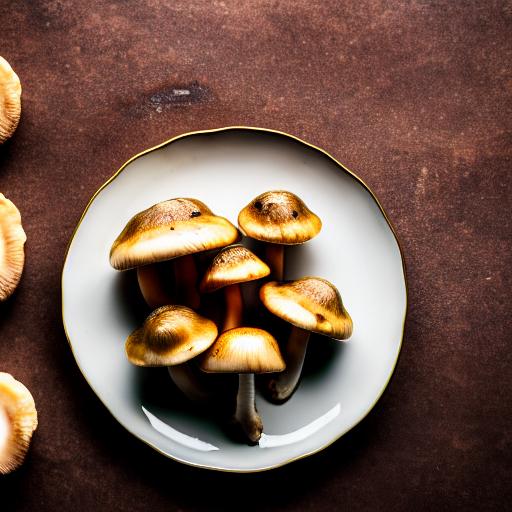}
    \includegraphics[width=1.5cm]{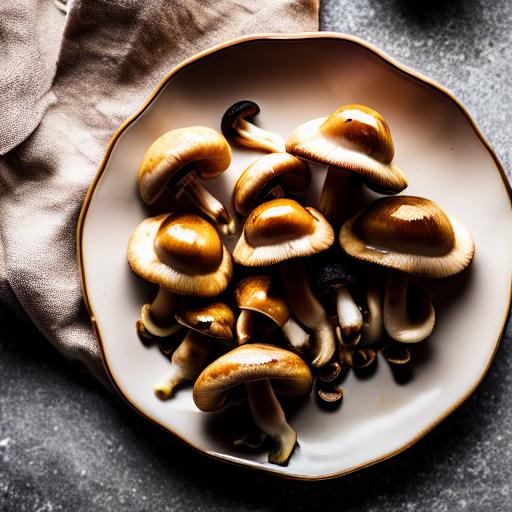}
    \caption{Chain produced using \sd and \textcaps.}
    \end{subfigure}\par\medskip
    \begin{subfigure}{\linewidth}
    \centering
    \includegraphics[width=1.5cm]{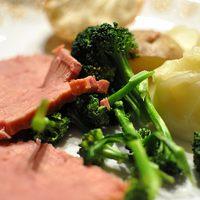}
    \includegraphics[width=1.5cm]{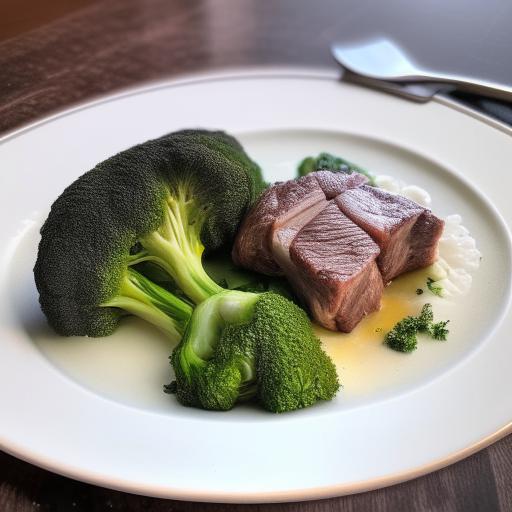}
    \includegraphics[width=1.5cm]{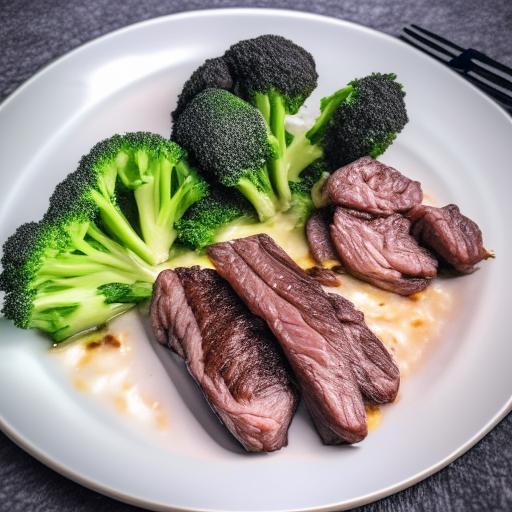}
    \includegraphics[width=1.5cm]{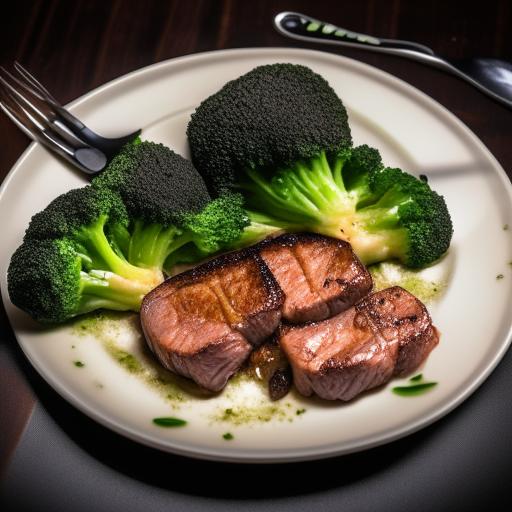}
    \includegraphics[width=1.5cm]{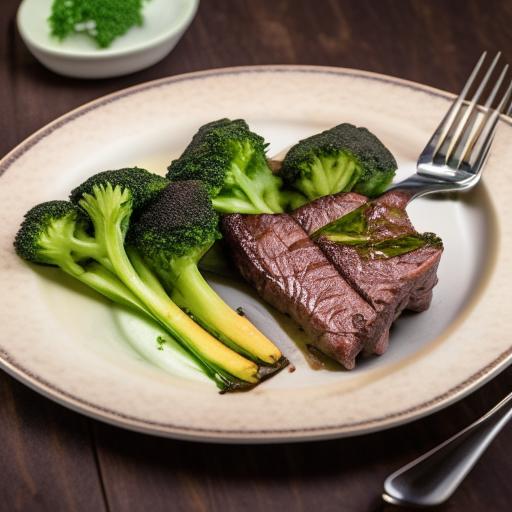}
    \includegraphics[width=1.5cm]{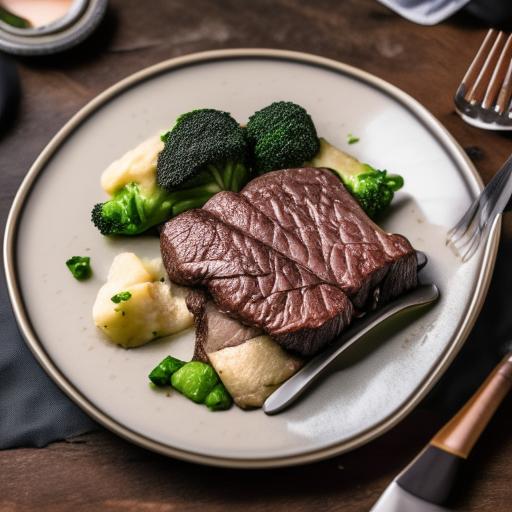}
    \includegraphics[width=1.5cm]{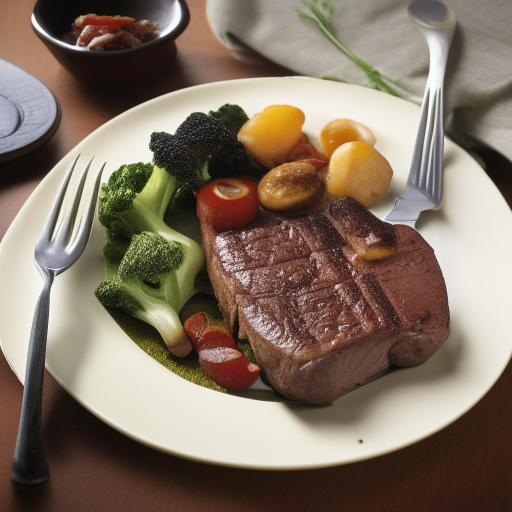}
    \includegraphics[width=1.5cm]{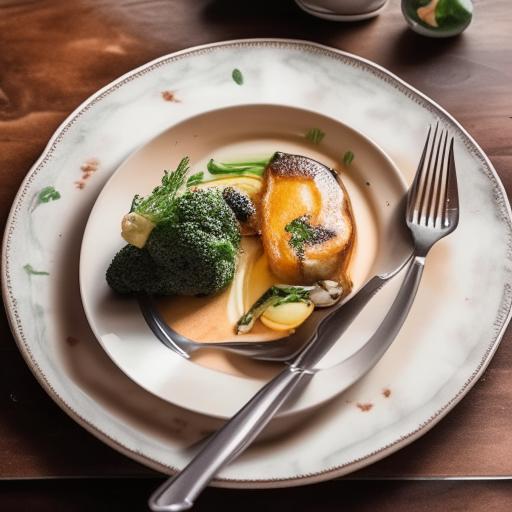}
    \includegraphics[width=1.5cm]{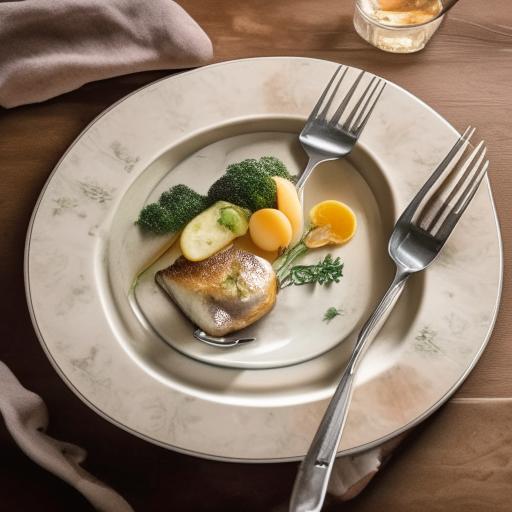}
    \includegraphics[width=1.5cm]{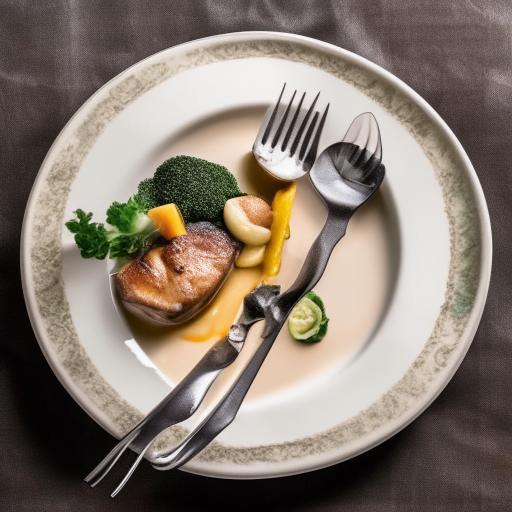}
    \includegraphics[width=1.5cm]{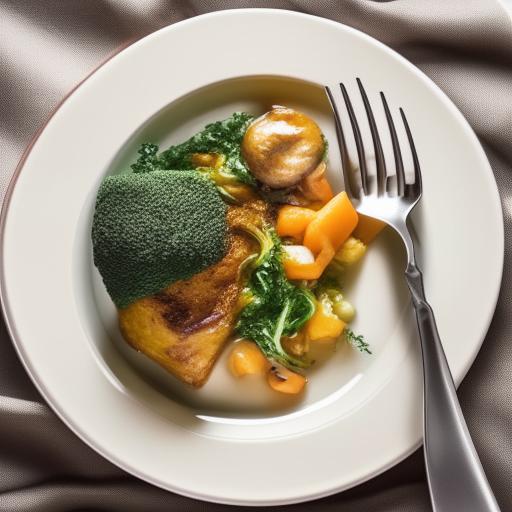}
    \includegraphics[width=1.5cm]{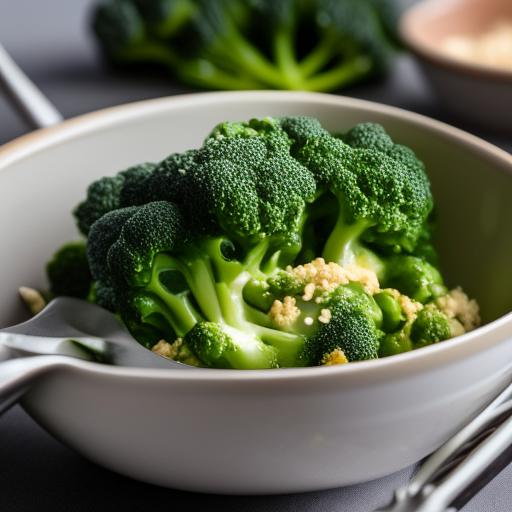}
    \includegraphics[width=1.5cm]{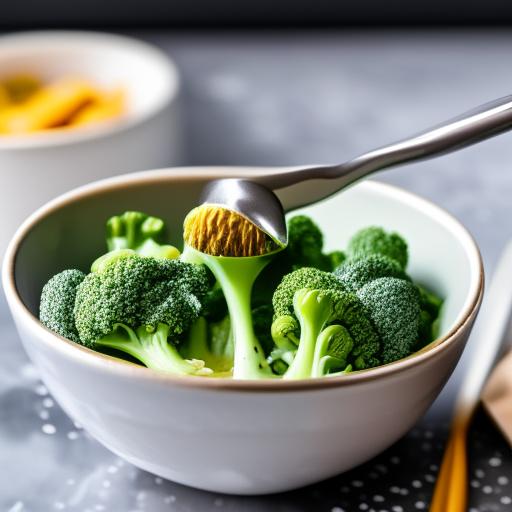}
    \includegraphics[width=1.5cm]{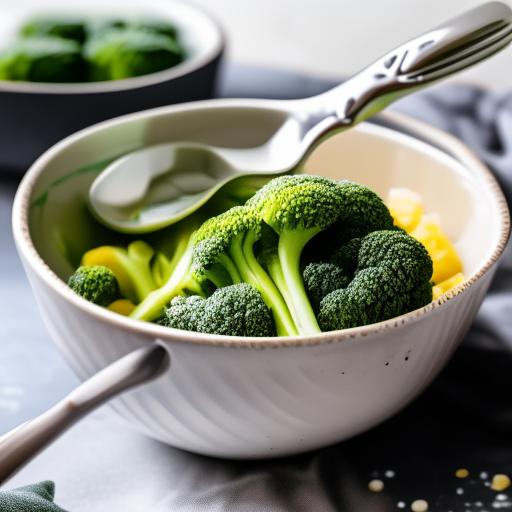}
    \includegraphics[width=1.5cm]{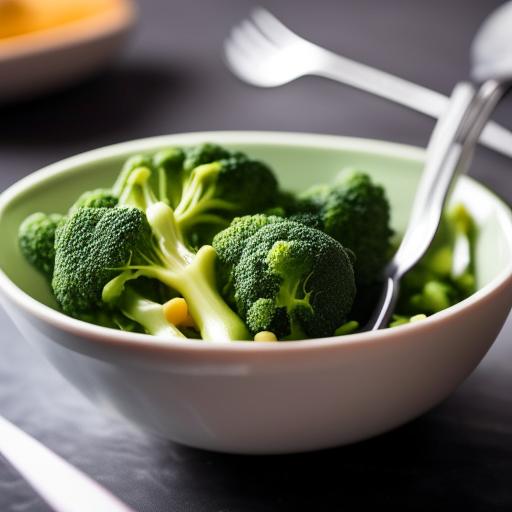}
    \includegraphics[width=1.5cm]{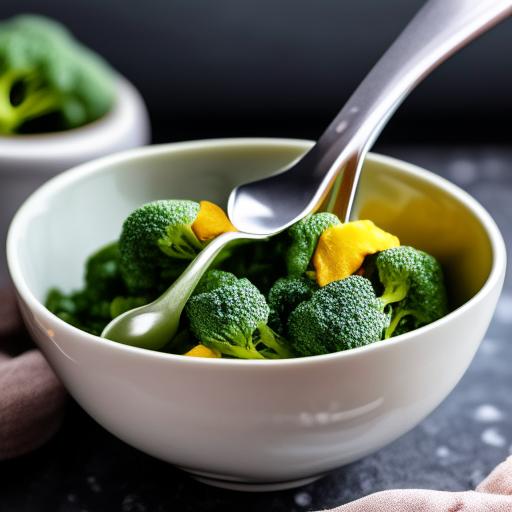}
    \caption{Chain produced using \kandinsky and \textcaps.}
    \end{subfigure}
    \caption{Example chains produced by ``ground truth'' image 0004, labeled ``broccoli'', where the top left is the ``ground truth'' image and the rest are generated.}
    \label{fig:chains-0004}
\end{figure}

\Cref{fig:chains-0004} displays a similar comparison between \kandinsky and \sd, this time captioned with \textcaps, with the former chain breaking after the 12th generated image, and the latter after the 2nd generated image. While the \kandinsky chain mostly kept the focus on the broccoli, only shifting to focus on the bowl at the end, the \sd chain adds a new element, mushrooms, and quickly makes that the focus of the chain. In these instances, \sd falls further toward fluidity on the faithful $\rightarrow$ fluid scale, while \kandinsky falls closer to faithful, a fact supported by their KL values.

The specific type of fluidity exhibited in these example chains aligns with the ``incorrect attribute binding'' glitch phenomenon defined in section 3.2 \citep{Chefer_2023}.

\begin{figure}
\centering
\begin{subfigure}{0.4\textwidth}
    \centering
    \includegraphics[width=0.5\textwidth]{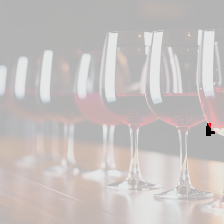}
    \caption{The 9th generated image, captioned ``several glasses of wine are lined up on a bar''}
    \label{fig:wine}
\end{subfigure}
\hfill
\begin{subfigure}{0.4\textwidth}
    \includegraphics[width=0.5\textwidth]{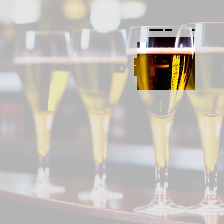}
    \caption{The 10th generated image, captioned ``four glasses of beer are lined up on a bar''}
    \label{fig:beer}
\end{subfigure}
\caption{Two images from a chain created from image 0071 using \sd and \blip: the image in \Cref{fig:beer} caused the chain to break, and \Cref{fig:wine} is the image immediately before it.}
\label{fig:rex-imgs}
\end{figure}

Other examples of chain breaks are harder to understand. Unlike the clear patterns of meandering depicted in the chains described earlier, the chain produced by ``ground truth'' image 0071, using a combination of \blip and \sd abruptly shifts from ``wine'' to ``bar'' between generated images 9 and 10, informing the rest of the chain, which then focuses on beer. At a glance, the 9th and 10th generated images (shown in \Cref{fig:rex-imgs}), between which the chain breaks, both looks like wine, so the low \clip and \yoloworld similarity score is difficult to understand. However, by using the ResNet152 image classification model and the \rex image explanation tool, the reason becomes more clear: the label given for the 9th generated image by ResNet152, from which the \rex explanation was created, is ``wine'', and a tiny portion of a wineglass highlighted as a sufficient visual explanation \citep{He_2016}. These models recognize ``beer'' in the 10th image, but produce a much larger visual explanation. This size discrepancy could indicate that the underlying structure of image 10 is more ambiguous than of image 9, requiring an object detector to take into account a larger area of pixels within the image structure in order to determine the most appropriate label. Building on this, one explanation for why some image generators are more creative than others can be surmised: a lack of global cohesiveness within a generated image due to the generation process could produce results that can straddle multiple class labels and cause shifts in prompt interpretation.

\section{Limitations}

Our approach comes with a number of limitations. Firstly, the code is quite expensive to run, as it utilizes multiple large models and generates roughly 15000 images per combination. Even using multiple GPUs and not rerunning finished chains, some of the slowest combinations can take over a day to run. In the future, we could optimize our code to resolve this issue. Our dummy chain construction was also constrained to just three common classes found in coco\_1000, as finding images for this purpose was somewhat difficult due to a lack of dedicated real-image datasets with labels that match our seed images. Perhaps including more classes would have changed the control group results, and potentially our statistical results. However, given the similarities we observed between the captions produced for different real-world images of simple objects like the ones in our seed image dataset, we believe this is unlikely.

\begin{figure}
\centering
\begin{subfigure}{0.4\textwidth}
    \centering
    \includegraphics[width=0.5\textwidth]{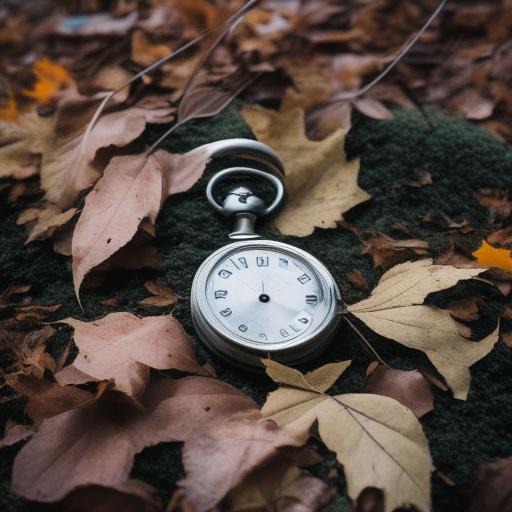}
    \caption{The 13th image in the chain.}
    \label{fig:leaves}
\end{subfigure}
\hfill
\begin{subfigure}{0.4\textwidth}
    \includegraphics[width=0.5\textwidth]{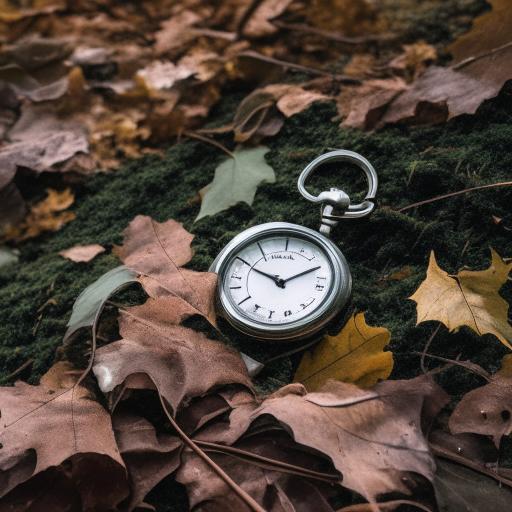}
    \caption{The 14th image in the chain.}
    \label{fig:moss}
\end{subfigure}
\caption{Two images from a chain created from image 0071 using \sd and \blip: the image in \Cref{fig:moss}, 
captioned ``a close up of a pocket watch on a moss covered ground'', caused the chain to break; \Cref{fig:leaves}, captioned ``there is a pocket watch sitting on the ground surrounded by leaves'', is the image immediately before it.}
\label{fig:8}
\end{figure}

Some chains produce breakage results which are difficult to understand: the \kandinsky \blip chain for ``ground truth'' image 0175 breaks between generated images 13 and 14, shown in \Cref{fig:8}, due to a low object detector similarity score, but despite having different labels ranked as ``most important'', both images are visually nearly identical. 
This makes it difficult to draw conclusions about some data produced by this experiment, although the larger statistical patterns still hold up under scrutiny.

Finally, it is important to acknowledge the impact machine creativity studies could have on human artists. There are already numerous instances of controversial ``co-creative'' collaborations between human and machine agents \citep{Daniele_2019}. Research which suggests that models may display autonomously creative behavior could be used in arguments against the importance of fully human-created art. This is not our intention, which is why we have emphasized equal, co-creative collaborations between human artists and AI.

\section{Conclusions}

We argue that glitches in image generation could be viewed not as mistakes, but as products of creative semantic interpretation, akin to how humans playing Chinese Whispers may come up with vastly different interpretations of given textual prompts. Our proposed new creativity measure, fluidity, can be quantified through using statistical analysis tools on a series of experiments which measure glitch frequencies in an array of image generators. Placing generators on our scale from maximally fluid to maximally faithful can allow users to choose a generation model based on creative behaviour, in accordance with their intended use case.
\newpage
\bibliographystyle{plainnat}
\bibliography{references,all}

\newpage
\appendix

\section{Technical Appendices}

\begin{algorithm}[h]
\caption{Pseudo-code for the LABEL\_SIM algorithm}
\label{fig:label-sim}
\begin{algorithmic}
\State $init\_img\_labels$
\State $curr\_img\_labels$
\State $similarity \gets 0$
\For{$l$ in $init\_img\_labels$}
    \If{$l$ in $m$}
        \State $similarity \gets similarity + 1$
    \Else
        \State $maxsim \gets 0$
        \For{each $l_2$ in $curr\_img\_labels$}
            \If{$maxsim \leq S-BERTScore(l, l_2)$}
                \State $maxsim \gets S-BERTScore(l, l_2)$
            \EndIf
        \EndFor
    \State $similarity \gets similarity + maxsim$
    \EndIf
\EndFor
\end{algorithmic}
\end{algorithm}

\begin{algorithm}
  \caption{Pseudo-code for our breakage algorithm}
  \label{fig:breakage-algorithm}
  \label{fig:1}
    \begin{algorithmic}
        \State $initial\_img$
        \State $current\_img$
        \State $initial\_img\_labels$
        \State $current\_img\_labels$
        \State $initial\_caption$
        \State $current\_caption$
        \State $breakage \ \gets \ False$
        \If{$CLIP_{score}(initial\_img, current\_img) \ < \ 20$}
            \State $breakage \ \gets \ True$
        \EndIf
        \If{${BERT_{score}(initial\_cap, current\_cap) < 0.5} \And\hfill\break {\sbert_{score}(initial\_cap, current\_cap) < 0.5}$}
            \State $breakage \ \gets \ True$
        \EndIf
        \If{$LABEL\_SIM(initial\_img\_labels, current\_img\_labels) \ < \ 0.5$}
            \State $breakage \ \gets \ True$
        \EndIf
    \end{algorithmic}
\end{algorithm}

Examples of experiments using \llava for image 0045 from the ground truth dataset: 
\begin{table}[!htb]
    \centering
    \includegraphics[scale=0.25]{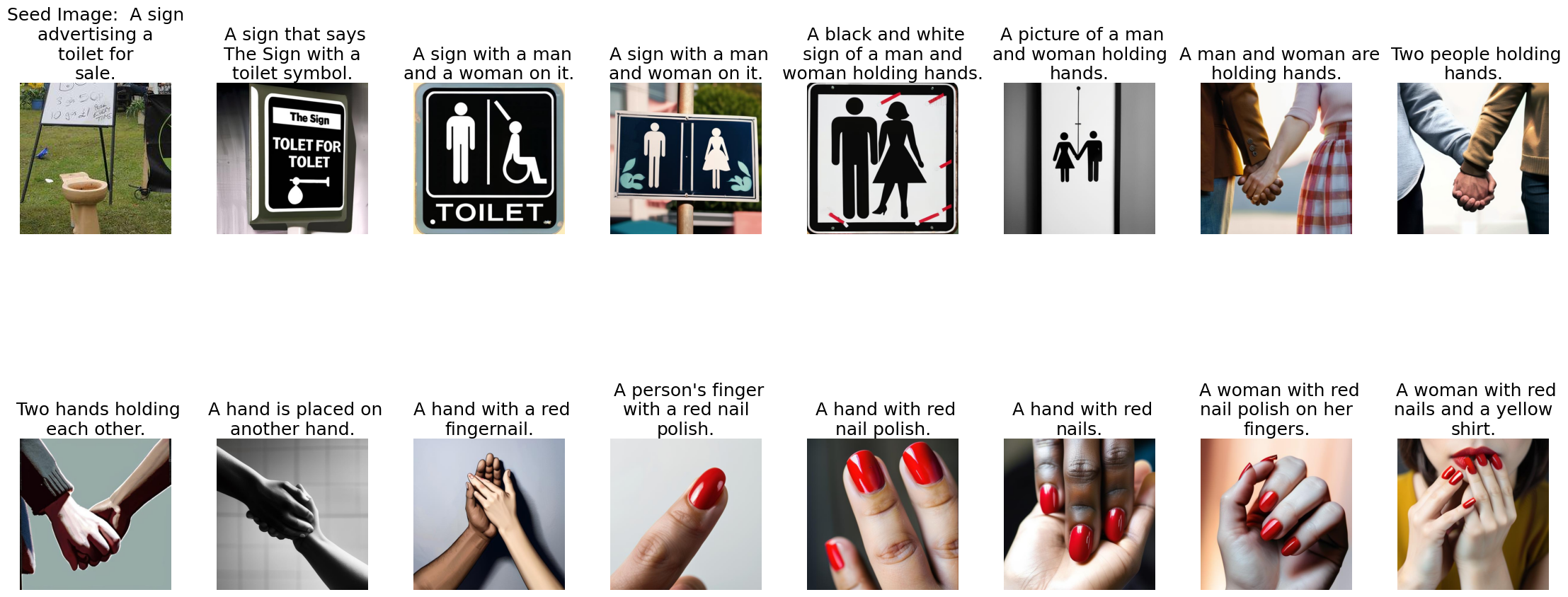}
    \caption{The chain constructed using \opendalle + \llava for image 0045}
    \label{fig:dalle-llava-appendix}
\end{table}

\begin{table}[!htb]
    \centering
    \includegraphics[scale=0.25]{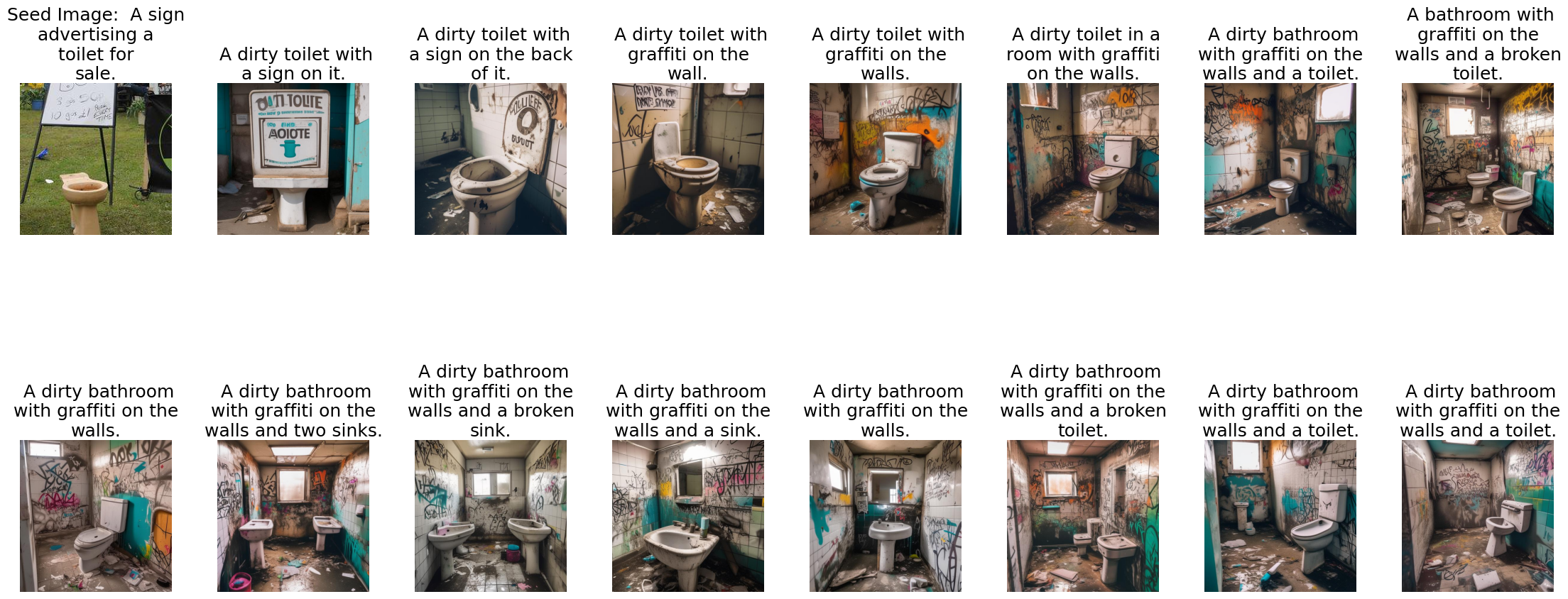}
    \caption{The chain constructed using \kandinsky + \llava for image 0045}
    \label{fig:kand-llava-appendix}
\end{table}

\begin{table}[!htb]
    \centering
    \includegraphics[scale=0.25]{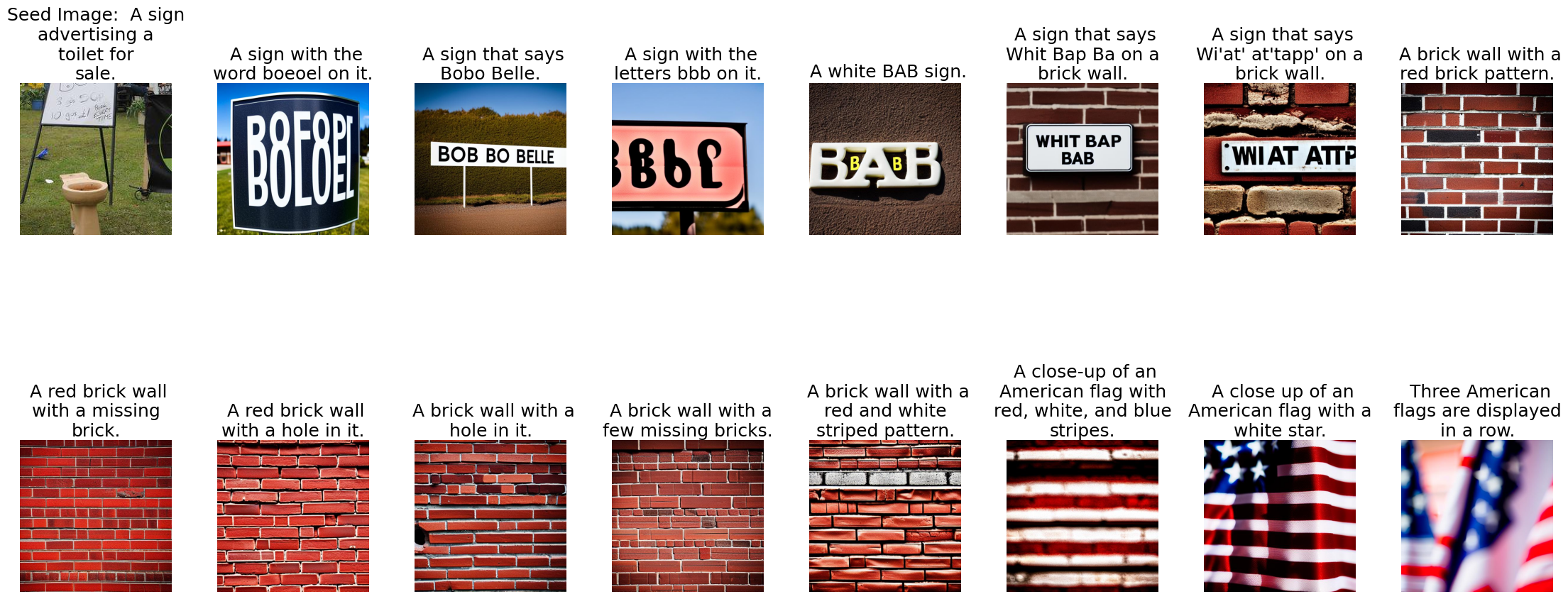}
    \caption{The chain constructed using \sd + \llava for image 0045}
    \label{fig:sd-llava-appendix}
\end{table}

\begin{table}[!htb]
    \centering
    \includegraphics[scale=0.25]{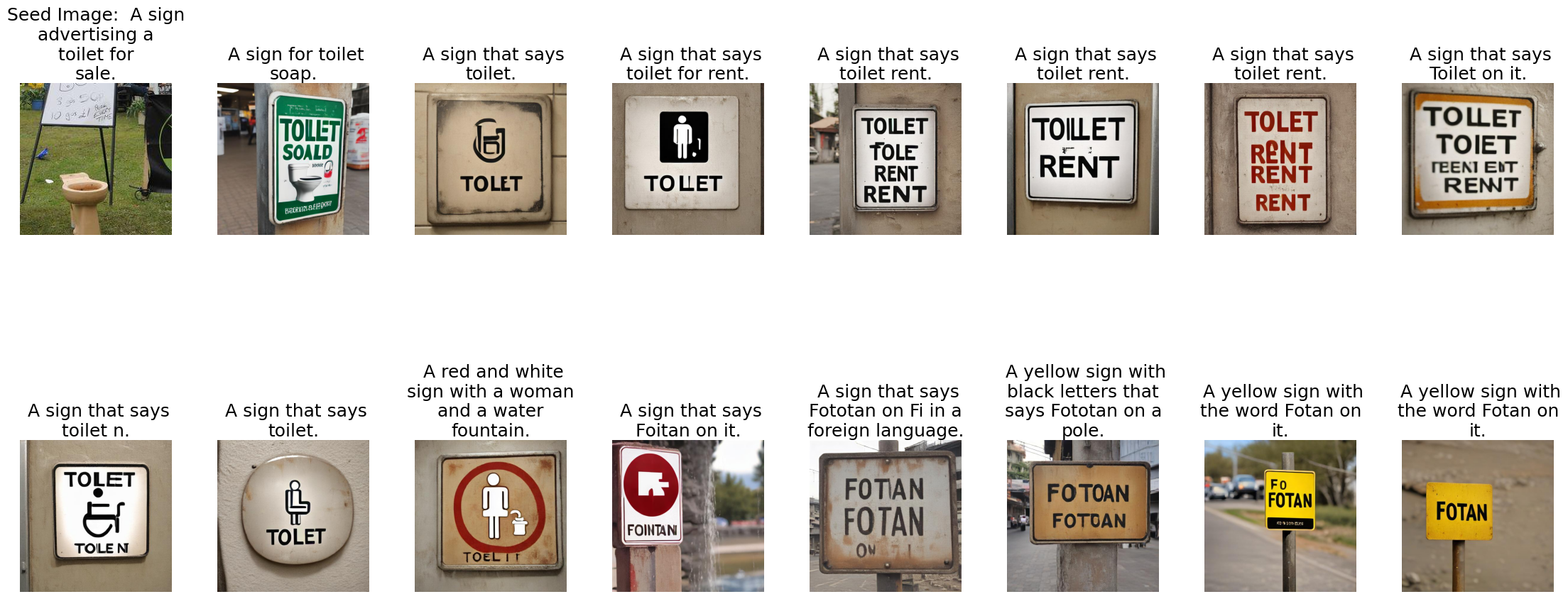}
    \caption{The chain constructed using \sdxl + \llava for image 0045}
    \label{fig:sdxl-llava-appendix}
\end{table}

The breakdown of the scores based on image and caption comparison for the chain shown in Figure \ref{fig:dalle-llava-appendix}.

\begin{table}[!htb]
\caption{The scores for image comparisons for img 0045 using \llava and \opendalle}
\label{fig:LabelScores}
    \centering
    \begin{tabular}{lp{1cm}lp{1cm}lp{1cm}lp{1cm}lp{1cm}lp{3cm}}
    \toprule
    & Clip Score & CLIP Label & Similarity of CLIP & YOLO Label  &  Similarity of YOLO & Broken  \\
\midrule
  0 & 21.7736 & ['sign advertising'] & 1     & ['sign advertising'] & 1     & False \\
  1 & 27.167  & ['toilet symbol']    & 0.123 & ['sign']             & 0.664 & False \\
  2 & 24.0346 & ['sign']             & 0.664 & []                  & 0     & False \\
  3 & 24.4452 & ['sign']             & 0.664 & []                  & 0     & False \\
  \arrayrulecolor{red}\midrule
  4 & 25.0439 & ['holding hands']    & 0.144 & []                  & 0     & True  \\
  \arrayrulecolor{red}\midrule
  5 & 23.9566 & ['holding hands']    & 0.144 & []                  & 0     & True  \\
  6 & 23.6537 & ['holding hands']    & 0.144 & []                  & 0     & True  \\
  7 & 24.1034 & ['holding hands']    & 0.144 & []                  & 0     & True  \\
  8 & 22.6222 & ['hands holding']    & 0.133 & ['person']           & 0.125 & True  \\
  9 & 23.9779 & ['hand']             & 0.126 & ['person']           & 0.125 & True  \\
 10 & 23.088  & ['hand']             & 0.126 & ['person']           & 0.125 & True  \\
 11 & 23.1896 & ['nail polish']      & 0.113 & ['person']           & 0.125 & True  \\
 12 & 21.3481 & ['nail polish']      & 0.113 & ['person']           & 0.125 & True  \\
 13 & 20.1922 & ['red nails']        & 0.085 & ['person']           & 0.125 & True  \\
 14 & 20.9572 & ['red nail']         & 0.045 & ['person']           & 0.125 & True  \\
 15 & 18.894  & ['red']              & 0.091 & ['person']           & 0.125 & True  \\
\arrayrulecolor{black}\midrule
\end{tabular}
\end{table}

\begin{table}
    \caption{The scores for initial and current caption textual comparisons for img 0045, using \llava and \opendalle}
    \label{fig:CaptionScores}
    \centering
\begin{tabular}{lp{5cm}lp{2cm}p{2cm}p{2cm}p{1cm}}
    & Caption & CLIP Score & BERT Score &   S-BERT Score &  Broken \\
\toprule
  0 & A sign advertising a toilet for sale.                    & 21.7736 & 1     & 1     & False \\
  1 & A sign that says The Sign with a toilet symbol.          & 27.167  & 0.762 & 0.789 & False \\
  2 & A sign with a man and a woman on it.                     & 24.0346 & 0.679 & 0.431 & False \\
  3 & A sign with a man and woman on it.                       & 24.4452 & 0.677 & 0.447 & False \\
  4 & A black and white sign of a man and woman holding hands. & 25.0439 & 0.621 & 0.248 & True  \\
  5 & A picture of a man and woman holding hands.              & 23.9566 & 0.615 & 0.108 & True  \\
  6 & A man and woman are holding hands.                       & 23.6537 & 0.573 & 0.038 & True  \\
  7 & Two people holding hands.                                & 24.1034 & 0.598 & 0.102 & True  \\
  8 & Two hands holding each other.                            & 22.6222 & 0.538 & 0.104 & True  \\
  9 & A hand is placed on another hand.                        & 23.9779 & 0.542 & 0.177 & True  \\
 10 & A hand with a red fingernail.                            & 23.088  & 0.672 & 0.196 & True  \\
 11 & A person's finger with a red nail polish.                & 23.1896 & 0.63  & 0.145 & True  \\
 12 & A hand with red nail polish.                             & 21.3481 & 0.65  & 0.209 & True  \\
 13 & A hand with red nails.                                   & 20.1922 & 0.682 & 0.174 & True  \\
 14 & A woman with red nail polish on her fingers.             & 20.9572 & 0.61  & 0.053 & True  \\
 15 & A woman with red nails and a yellow shirt.               & 18.894  & 0.624 & 0.073 & True  \\
 \bottomrule
\end{tabular}
\end{table}

\begin{table}
\caption{Statistics for Frequency Distributions for Each Combination}
\label{tab:individual-stats}
    \centering
    \begin{tabular}{rllrrrr}
\toprule
Model& Caption   &   KL Divergence &   Mean Chain Length &   Skewness &P-Value \\
\midrule
\kandinsky & \blip& 0.711418 &   7.399 &    3.85383 & 0.000116283 \\
\kandinsky & \textcaps  & 0.575388 &   6.496 &    3.29603 & 0.000980611 \\
\kandinsky & \llava& 0.945868 &   8.776 &    4.41298 & 1.01959e-05 \\
\opendalle & \blip& 0.44288  &   6.728 &    3.43983 & 0.00058209  \\
\opendalle & \textcaps  & 0.381594 &   5.309 &    2.37619 & 0.0174923   \\
\opendalle & \llava& 0.576568 &   7.366 &    3.92607 & 8.63434e-05 \\
\sd & \blip& 0.410231 &   5.099 &    2.58584 & 0.00971408  \\
\sd & \textcaps  & 0.458674 &   4.384 &    2.82806 & 0.00468312 \\
\sd & \llava& 0.470352 &   5.679 &    2.69947 & 0.0069451   \\
\sdxl & \blip& 0.685981 &   7.949 &    4.24887 & 2.14849e-05 \\
\sdxl & \textcaps  & 0.603585 &   7.024 &    3.67724 & 0.000235768 \\
\sdxl & \llava& 0.78107  &   8.331 &    4.32471 & 1.52734e-05 \\
\bottomrule
\end{tabular}
\end{table}

\begin{table}
\caption{Statistics for Frequency Distributions for Control Combinations}
\label{tab:control-stats}
    \centering
    \begin{tabular}{rllrrrr}
\toprule
Model& Caption   &   KL Divergence &   Mean Chain Length &   Skewness &P-Value \\
\midrule
Control& \blip& 2.169674  &   13.523 &    4.90856 & 9.174705e-07  \\
Control& \textcaps  & 2.078700 &   12.777 &    4.86619 & 1.13773e-06   \\
Control& \llava& 2.180563 &   12.619 &    4.75709 & 1.96402e-05 \\
\bottomrule
\end{tabular}
\end{table}

\begin{table}
\caption{Mann-Whitney U Comparisons for Control + Caption Gen \& Image + Caption Gen Combinations}
\label{tab:control-comparisons}
    \centering
    \begin{tabular}{rllrrrr}
\toprule
Image Generator 1 & Caption Generator   &   p-value \\
\midrule
\opendalle & \blip & 0.0005 \\
\opendalle & \llava & 0.0004 \\
\opendalle & \textcaps & 0.0006 \\
\kandinsky & \blip & 0.0012 \\
\kandinsky & \llava & 0.0005 \\
\kandinsky & \textcaps & 0.0011 \\
\sd & \blip & 0.0005 \\
\sd & \llava & 0.0003 \\
\sd & \textcaps & 0.0009 \\
\sdxl & \blip & 0.0008 \\
\sdxl & \llava & 0.0004 \\
\sdxl & \textcaps & 0.0012 \\
\bottomrule
\end{tabular}
\end{table}

\begin{table}
\caption{Mann-Whitney U Comparisons for Image Generators}
\label{tab:img-comparisons}
    \centering
    \begin{tabular}{rllrrrr}
\toprule
Caption Generator & Image Generator 1 & Image Generator 2  &   p-value \\
\midrule
\blip & \kandinsky & \opendalle & 0.2538 \\
\blip & \kandinsky & \sd & 0.4426 \\
\blip & \kandinsky & \sdxl & 0.7086 \\
\blip & \sdxl & \opendalle & 0.5067 \\
\blip & \sdxl & \sd & 0.5894 \\ 
\blip & \sd & \opendalle & 0.7874 \\ 

\llava & \kandinsky & \opendalle & 0.2447 \\
\llava & \kandinsky & \sd & 0.2537 \\
\llava & \kandinsky & \sdxl & 0.4550 \\
\llava & \sdxl & \opendalle & 0.5609 \\
\llava & \sdxl & \sd & 0.5473 \\ 
\llava & \sd & \opendalle & 0.9503 \\ 

\textcaps & \kandinsky & \opendalle & 0.5894 \\
\textcaps & \kandinsky & \sd & 0.8845 \\
\textcaps & \kandinsky & \sdxl & 1.0 \\
\textcaps & \sdxl & \opendalle & 0.5066 \\
\textcaps & \sdxl & \sd & 0.8519 \\ 
\textcaps & \sd & \opendalle & 0.7398 \\ 
\bottomrule
\end{tabular}
\end{table}

\begin{table}
\caption{Mann-Whitney U Comparisons for Caption Generators}
\label{tab:cap-comparisons}
    \centering
    \begin{tabular}{rllrrrr}
\toprule
Image Generator & Caption Generator 1 & Caption Generator 2  &   p-value \\
\midrule
CONTROL & \blip & \llava & 0.8736 \\
CONTROL & \blip & \textcaps & 0.5180 \\
CONTROL & \llava & \textcaps & 0.6891 \\

\opendalle & \blip & \llava & 0.5753 \\
\opendalle & \blip & \textcaps & 0.9504 \\
\opendalle & \llava & \textcaps & 0.7872 \\

\kandinsky & \blip & \llava & 0.5335 \\
\kandinsky & \blip & \textcaps & 0.8194 \\
\kandinsky & \llava & \textcaps & 0.4426 \\

\sd & \blip & \llava & 0.8194 \\
\sd & \blip & \textcaps & 0.8194 \\
\sd & \llava & \textcaps & 1.0 \\

\sdxl & \blip & \llava & 0.8033 \\
\sdxl & \blip & \textcaps & 0.9338 \\
\sdxl & \llava & \textcaps & 0.6781 \\
\bottomrule
\end{tabular}
\end{table}


\end{document}